%% file: sample-manuscript.tex
\documentclass[sigconf]{acmart}
\usepackage[english]{babel}
\usepackage{placeins}
\usepackage{textcomp}
\usepackage{graphicx}
\usepackage{xcolor}
\usepackage[normalem]{ulem}
\usepackage{siunitx}
\copyrightyear{2026}
\acmYear{2026}
\setcopyright{cc}
\setcctype{by-nc-nd}
\acmConference[C\&C '26]{Creativity and Cognition}{July 13--16, 2026}{London, United Kingdom}
\acmBooktitle{Creativity and Cognition (C\&C '26), July 13--16, 2026, London, United Kingdom}
\acmDOI{10.1145/3803784.3807547}
\acmISBN{979-8-4007-2583-8/2026/07}

\AtBeginDocument{%
  }

\newcommand{\new}[1]{\textcolor{black}{#1}}

\newcommand{\newcc}[1]{\textcolor{black}{#1}}

\begin{document}

\title{Art Card Game (ACG): Embedding Illustration in Gameplay to Mitigate Artist Self-Criticism}

\author{Catherine Mullings}
\email{cmulling@stanford.edu}
\orcid{0009-0008-9892-4768}
\affiliation{%
  \institution{Stanford University}
  \city{Stanford}
  \state{California}
  \country{USA}
}

\author{Michael S. Bernstein}
\email{msb@cs.stanford.edu}
\orcid{0000-0001-8020-9434}
\affiliation{%
   \institution{Stanford University}
  \city{Stanford}
  \state{California}
  \country{USA}
}



\begin{abstract}
\newcc{Persistent self-criticism---harsh evaluative self-talk---can undermine illustrators’ performance and well-being. Traditional interventions draw on psychotherapeutic approaches (e.g., compassion training) but sit outside the illustration workflow, requiring time, facilitation, and skill transfer. We propose an in-workflow alternative: \emph{evaluative off-centering}, a mechanism redirecting self-critical evaluation away from an inherently self-evaluative task (like illustration) by embedding it in an alternative activity. We instantiate evaluative off-centering in Art Card Game (ACG) that integrates illustration into a card customization game: players illustrate cards that become playable assets in a head-to-head battle. In a four-day randomized controlled study with hobbyist and professional illustrators (\mbox{$N{=}38$}), ACG outperformed a control condition with identical illustration constraints but no evaluative off-centering mechanisms (e.g. multiplayer, gameplay), yielding significantly higher pride in produced artwork and activity enjoyment. Pride and enjoyment---positive affect states linked to lower self-criticism---help explain how ACG reduces self-criticism. We discuss design implications for creativity support tools that apply evaluative off-centering across creative domains.}
\end{abstract}

\ccsdesc[500]{Human-centered computing~Collaborative and social computing systems and tools}

\keywords{creativity support tools, self-criticism, games}

\begin{teaserfigure}
    \begin{minipage}{\textwidth}
      \includegraphics[width=\textwidth]{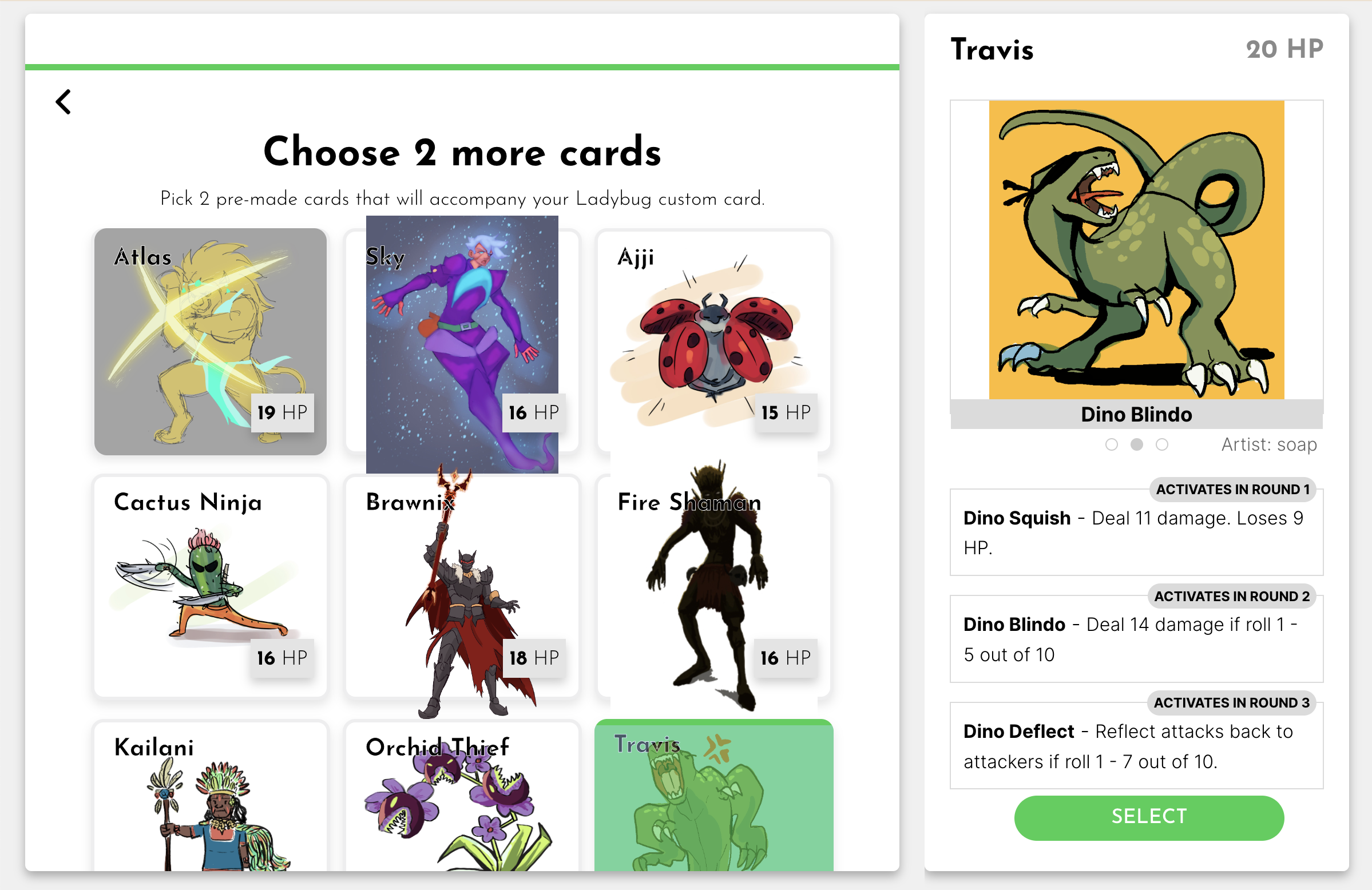}
    \end{minipage}
    \caption{We present an illustration game, Art Card Game (``ACG''), where players create and draw character cards as they play a competitive card game. By focusing the illustrator's attention on the game, ACG helps mitigate self-criticism during artmaking.}
    \label{fig:flow}
    \Description{The Art Card Game interface displaying the deck card selection screen. The interface is divided into a larger left panel containing a grid of options and a narrower right panel showing a preview with details of the currently selected option. The left panel shows nine premade deck cards, each showing an illustration, character name, and hit-point value. One card at the bottom is highlighted, indicating it is selected. The right panel previews more of the selected card’s details, like the illustration (larger size), the illustrator’s username, and the three moves associated with that card, each labeled with the move name, description of the move’s effect, and activation round. In this particular screenshot, the selected card features artwork of a blind-folded Tyrannosaurus rex roaring. Below the artwork is the associated move name (``Dino Blindo'').}
\end{teaserfigure}

\maketitle

\section{Introduction}
Illustrations saturate contemporary life---shaping brands, interfaces, and entertainment across print and screen---and serve as a primary channel for personal expression~\cite{whatisillustration}. However, producing illustration is cognitively and affectively demanding~\cite{KochJulieM2024FASM, carson2011}, and is frequently accompanied by persistent self-criticism among hobbyist and professional illustrators~\cite{standish2022}. For illustrators, self-criticism---the tendency to harshly and punitively judge and scrutinize oneself~\cite{shahar2012pilot}---can disrupt one’s illustration performance and lead to lengthy periods of revision~\cite{SION2023102060}; reduce the generation of creative ideas~\cite{GLAZIEWICZ2024101692}, induce “excessive sensitivity to mistakes”~\cite{BASAK20125010}, and contribute to burnout~\cite{GLAZIEWICZ2024101692}. Despite these adverse effects, illustrators often justify persistent self-criticism as a means to perform well~\cite{whatisillustration}. How might we help illustrators reduce self-criticism during artmaking enough that they can finish work and gain experience as they learn how to manage the critic on their shoulder?

Prior work has explored methods to reduce self-criticism, but they are typically delivered as standalone interventions rather than integrated into the creative workflow. Skills-based approaches---compassion training~\cite{Germer2013SelfcompassionIC}, CBT/ACT for perfectionism~\cite{galloway2022efficacy, ong2019randomized}, and dialogue-based therapies~\cite{shahar2012pilot}---are delivered via coached sessions and self-help media (books, videos), with some writing exercises~\cite{zabelina2010don}. While promising, these approaches provide generic guidance, require dedicated time or facilitation, and seldom target the immediate, task-specific decisions of illustration. Motivational art videos~\cite{chiu2016,  adamduff2020} can be played while working, but advice is typically not tied to the artifact at hand, and their long\new{-}term effectiveness is not well studied. Peer and social support delivered via forums~\cite{r-artistlounge} can be beneficial\new{,} yet are often after-the-fact, dependent on who is present, and may increase evaluation apprehension. Accordingly, there is a need for in-situ support that operates during artmaking. These approaches face adoption and fit constraints for illustrators (time, facilitation, mixed evidence, and limited validation in studio workflows), motivating in-situ alternatives.

\newcc{We pursue a strategy that we call \emph{evaluative off-centering}: redirecting self-critical evaluation away from a focal task that is prone to self-judgment by embedding it in an alternative, surrounding activity that is intrinsically motivating. Unlike traditional interventions to self-criticism that are separate from the creative workflow, evaluative off-centering is an in-workflow alternative: illustrators can reap the benefits of reduced self-criticism \emph{while} \new{artmaking}. This is achieved by prioritizing the \new{surrounding activity's goals}, which occupies the artist's evaluative attention while ensuring the primary creative task is still completed.}

\newcc{Our intuition to apply evaluative off-centering to the creative process is grounded in evaluative off-centering's effectiveness across various domains where performance anxiety or self-consciousness typically hinder engagement. For example, karaoke prioritizes social participation over vocal accuracy~\cite{bryan2025playing}; \emph{Zombies, Run!} reframes exercise as narrative survival rather than a fitness test~\cite{ZRun}; and \emph{Human Resource Machine} frames learning programming as office puzzle games rather than competence tests~\cite{hrm}. In each case, the surrounding activity shifts focus from self-evaluation to immediate, achievable objectives.} 
 
\newcc{Given the efficacy of evaluative off-centering in other domains, we see its potential to address the self-criticism inherent in visual illustration. To leverage this mechanism, we must identify a surrounding activity that diverts focus from aesthetic quality. While there are many possible activities, we pursue a game-based approach because it conveniently packages several key evaluative off-centering properties that are capable of capturing and holding illustrators' attention: 1) social interaction, 2) art-as-a-ticket to entry (creating art as input that unlocks a game's progress; progression is not dependent on the art quality), and 3) strategic play. To this end, we embody these properties and instantiate evaluative off-centering in \textbf{Art Card Game (ACG)}, a two-player online game that integrates illustration into a card customization workflow: players illustrate cards that become immediately playable assets in a head-to-head battle.} By binding image-making to concrete, time-bounded goals and prompts, ACG makes artmaking \emph{part of the game loop rather than the sole object\new{ive} of the game}, shifting attention from ``make this perfect'' toward ``make this work in gameplay'' and offering relief from momentary self-criticism.

We evaluate ACG in a four-day randomized experiment with illustrators (\mbox{$N{=}38$}), contrasting the (1)~ACG experimental condition against (2)~a control condition that preserves the same character drawing prompts, the same move$\rightarrow$pose binding, and identical time constraints but is solo and has no  game. The experimental condition yielded a \emph{reliable increase in pride} in one’s \new{artwork} and a \emph{tendency toward higher enjoyment}. Pride and enjoyment are positive-affect~\cite{watson1988development} states closely tied to lower momentary self-criticism~\cite{zuroff2016conceptualizing}. Qualitative responses---surveys and post-study interviews---illuminated why these effects emerged\newcc{. Across our findings, evaluative off-centering showed up in two related outcomes: participants felt better about (1) their resulting work-in-progress (e.g., pride and ownership toward rough artifacts) and/or (2) the illustration process (e.g., confidence and willingness to experiment).}

This paper contributes: \emph{(1)} the concept of \textbf{evaluative off-center--
ing}---\newcc{embedding a creative task in an alternative, surrounding activity that is intrinsically motivating}---as an approach for reducing momentary self-criticism during creative work; \emph{(2)} the design and implementation of \textbf{Art Card Game}, \newcc{which instantiates evaluative off-centering in a socially situated, strategy-driven card game where illustrations are required for participation and immediately used in play}; and \emph{(3)} a mixed-methods \textbf{evaluation} showing affective benefits relative to a time-matched solo baseline and explaining \emph{why} they arise. We close by outlining design implications for creativity-support tools.

\section{Related Work}
In this section, we provide an overview of the existing literature on self-criticism and its remedies, followed by a survey of art games in relation to ACG. Throughout, we will highlight key areas where this research makes a contribution.

\subsection{Self-Criticism in Creative Work}
Self-criticism---the tendency to harshly  judge and scrutinize oneself~\cite{shahar2012pilot}---is an all-too-familiar companion for illustrators. In a domain where high-quality results are praised~\cite{kim2017mosaic}, illustrators face a high expectation to produce high\new{-}quality, finished work. A closer look into the illustration workflow---with all its many interdependent stages (e.g., composition, linework, color, rendering) and continual artistic decisions\footnote{``Drawing is not a walk in the park, its so stressful and takes a ton of brain power between composition, line quality, color, subject matter and everything else.''---r/ArtistLounge Reddit User 1~\cite{redditPerfection1}}---helps explain why revisions, feelings of distress, and ultimately harsh self-criticism are common\footnote{``I'm constantly finding mistakes in everything I do, keep making touch-ups in the sketch process, and just imagining how some people may react negatively to it.''--r/ArtistLounge Reddit User 2~\cite{redditPerfection1}}. The effects of self-criticism on illustrators are striking: disrupting one’s illustration performance and leading to lengthy periods of revision~\cite{SION2023102060}; reducing the generation of creative ideas~\cite{GLAZIEWICZ2024101692}, inducing ``excessive sensitivity to mistakes''~\cite{BASAK20125010}, and contributing to burnout~\cite{GLAZIEWICZ2024101692}. In this section, we review existing solutions to mitigate self-criticism in creative work.

\paragraph{Creativity Tools for Alleviating Self-Criticism}
Beyond clinical interventions for self-criticism---such as self-compassion training~\cite{Germer2013SelfcompassionIC} and CBT/ACT~\cite{galloway2022efficacy, ong2019randomized}---that are typically practiced \emph{outside} the act of illustration and require illustrators to translate generic guidance to artistic practice, there exists HCI research on creativity-support tools focused on improving artists' affect like reducing self-criticism or increasing motivation \emph{during} creative flow~\cite{sternam2022}. SonAmi is a creative writing tool that allows writers to listen and reflect on audio transcriptions of their written text to reduce self-criticism that demotivates writers and leads \new{to} procrastination~\cite{belakova2021}. Although its underlying mechanism of \emph{creative voicing} helped writers edit, revise, and gain new perspective and ideas without losing focus or becoming discouraged, it is likely more suitable for written tasks than for illustrative ones. Applying it to illustration would require transcribing in-progress images to text, which alters the original input (the image) and can interrupt the illustration flow. 

Closer to the domain of illustration, Mosaic was an art feedback, online community that normalized sharing works-in-progress pieces over polished, finished artworks, encouraging ``useful creative outcomes---such as mistakes, failures, prototypes, and experiments''~\cite{kim2017mosaic}. Although works-in-progress were deemed as shareable and valued, Mosaic made illustration the focal point, whereby some users reported instances of self-criticism (as one Mosaic user put it, ``You start doing something and it looks horrible'') and apprehension to share their work. ACG builds upon Mosaic’s process-first ethos, but shifts the underlying framing of illustration: rather than treating illustrations as the focal point for feedback and aesthetic evaluation, ACG \newcc{off-centers the illustration process and redirects attention to gameplay}. Our contribution compares this \newcc{\emph{evaluative off-centering}} approach with an illustration-centered baseline and evaluates its effects on self-criticism measures.

Another application of process-first interventions in creative work is the technique of \emph{parallel prototyping} from design, which asks designers to first explore and quickly prototype multiple solutions to a problem rather than committing and refining a single solution. This approach has been shown to lead to more effective solutions, divergent ideas, and increases designers' receptivity to critique~\cite{dow2010}. Illustrators employ an analogous problem solving technique called \emph{thumbnail sketching} to rapidly sketch and explore multiple different ideas before settling on one~\cite{thumbnailing2023}. Although ACG targets the refinement instead of the ideation phase of illustration, we draw upon the same principals of parallel prototyping: illustrators are given a series of short time-constrained tasks that emphasize breadth over depth and encourage them to work quickly and efficiently to produce ``good enough'' outcomes.

\new{Similar to parallel prototyping’s goal of producing divergent outcomes, another process-first intervention is defamiliarization, which disconnects users from their habitual creative process so they can generate unexpected or strange results~\cite{carlson2013designing}. An example of defamiliarization in visual arts includes setting constraints in material choice (e.g., only use wood) to create novel results~\cite{noguchi1999material}. However, while defamiliarization may be effective for provoking innovation, it intentionally requires users to problem-solve in an unfamiliar context~\cite{crawford1984viktor, carlson2013designing}, making the creative task more demanding and disorienting~\cite{carlson2013designing}. In highly disorienting configurations, users may be ``unable to move forward on their own''~\cite{carlson2013designing}. While ACG incorporates elements adjacent to defamiliarization (time constraints, card game context), our primary goal is not to heighten disorientation to create novelty, but to off-center attention away from self-evaluation during creative workflows.}

\subsection{\newcc{Evaluative Off-centering}}
While many creativity support tools promote process-first practices and efficient exploration, they still keep the creative process itself as the focal activity---which can keep evaluative attention centered on the work-in-progress and trigger self-criticism that disrupts making. \new{One strategy for mitigating this is to redirect the artist’s focus away from their own performance. In the concept of locus of control, focus can be shifted toward evaluating the performance of an external source (external locus of control)~\cite{lefcourt1991locus}. For example, instead of novice programmers blaming themselves for programming errors, in Gidget~\cite{leeko2011}, the blame is transferred to the tool itself.}

\new{Shifting the locus of control and redirecting blame to an external source (like a programming tool) may not be the most effective strategy for creative work, where ``errors'' are often subjective, ambiguous, and deeply tied to the creator's internal sensitivity.} \new{Instead} we pursue \emph{evaluative off-centering}: redirecting evaluative attention away from a self-evaluative focal task by embedding it within an intrinsically motivating surrounding activity. \new{We introduce this term to give name to a recurring design pattern observable in other domains; in this paper, we explore off-centering's applicability to creative workflows.} \new{Note that, o}ff-centering targets tasks hindered by internal judgment, distinct from interventions for rote habits (e.g., incentivizing toothbrushing via \textit{Pokémon Smile}~\cite{hwang2023effect}). Below, we synthesize examples across domains and highlight design elements---notably social interaction, narrative immersion, and gameplay---that shift attention away from evaluative tasks.

\subsubsection{\newcc{Evaluative off-centering via Social Framing}}
\newcc{Social interaction in a surrounding activity can support evaluative off-centering by shifting attention from solitary self-monitoring toward shared experience, norms, and roles. Karaoke, for instance, off-centers singing by embedding it within a social activity among friends that rewards participation and shared fun rather than vocal quality, often with peers singing in unison to reduce the spotlight on any one performer~\cite{bryan2025playing}. Digital systems can similarly leverage social roles through teachable agents: students teach a character whose performance reflects what they have taught, shifting attention from self-evaluating mistakes to caring about and revising what the agent knows~\cite{2009JSEdT,reeves1996media}.}

\subsubsection{\newcc{Evaluative off-centering via Narrative Immersion}}
\newcc{Narratives can off-center self-evaluative tasks by linking engagement to story progression, making the activity feel like participation in an unfolding experience rather than a performance to judge. In Health-HCI, narrative immersion is often used to promote exercise in contexts where people may experience social physique anxiety---anxiety about others observing or evaluating one’s body~\cite{hart1989tie}. Applications like \emph{Zombies, Run!}~\cite{ZRun} and \emph{WhoIsZuki}~\cite{murnane2020designing} embed movement within an immersive narrative, so users are primarily advancing the story (e.g., surviving a threat), with exercise functioning as the ticket that unlocks what happens next.}

\subsubsection{\newcc{Evaluative off-centering via Gameplay}}
\newcc{Gameplay can off-center a self-evaluative task by redirecting attention to the immediate demands of play---responding, choosing, and adapting to achieve in-game goals. This is particularly robust for skills associated with performance anxiety: \emph{The Typing of the Dead}~\cite{ToTD} off-centers typing pressure by mapping accuracy directly to combat mechanics. A similar strategy appears in programming-oriented games: because repeated negative feedback (e.g., code that fails to compile) can heighten anxiety~\cite{4569871}, \emph{Human Resource Machine} reframes ``writing code'' as manipulating concrete box-moving operations in an office puzzle, shifting attention toward step-by-step actions and outcomes rather than fear of failure~\cite{hrm}.}

\subsection{Games for Art}
In designing ACG, we surveyed existing art games to understand what underlying mechanisms make those games enjoyable, where they fall short in reducing illustrators' self-criticism, and what mechanisms we \new{adopt}.

\paragraph{Party Art Games and Time Constraints} Our work draws inspiration from art party games such as Gartic Phone~\cite{GRTPHNE} and Pictionary, which put time constraints on a group art activity. Time constraints in games are used to develop a sense of urgency and challenge players’ abilities to respond under pressure. In Pictionary-esque games, the artwork is discarded each round, encouraging less effort and lower quality artwork \new{like} doodles. In contrast, ACG aims to straddle a design space between untimed illustration\footnote{\new{Generally}, artists spend several hours or days completing a\new{n} illustration.} and rapid throwaway art. We leverage time constraints to reduce perfectionism, but because the artwork persists as a playable card, it encourages more effort than disposable doodles.

\paragraph{Online Art Challenges and Artistic Prompts}
Online art challenges use themed prompts to deliver comparable creative rewards and social engagement as games. Artistic prompts act as guides to help illustrators overcome \emph{art block}~\cite{pressfield2002} by narrowing the problem space and jump-starting ideas. The Inktober challenge, for example, releases daily prompts during October, and illustrators post their daily creations on social media to showcase participation~\cite{InkTober}. In ACG, we adopt a similar prompt-based approach: we provide rich character descriptions that give players the structure, yet freedom to create. Moreover, like in an online art challenge, ACG players have the opportunity to view others' submissions, exposing them to alternative prompt interpretations and styles. 

\paragraph{Art Trading Games and Character Art}
Online art-trading communities revolve around the strong sense of ownership artists feel toward their original characters (OCs). For example, \emph{Art Fight}~\cite{af} exemplifies the OC-trading model: participants ``attack'' by drawing other players' OCs, earning points for complexity and execution. However, \emph{Art Fight}'s competitive cadence can incentivize high-effort pieces on tight timelines and contribute to burnout.\footnote{For discussion of burnout, see \url{https://sfw.furaffinity.net/journal/10261042}.} Since character art is both popular and personally meaningful to artists~\cite{character2020}, ACG centers on character art, but excludes the art trading model to avoid backlogs of owed trades and pressure to polish work. 

\paragraph{Games with positive psychological outcomes} Many art games are solely focused on fun while others are aimed at practicing or learning technical artistic skills~\cite{williford2017zensketch}, but none aims to address the meta-cognitive or psychological outcomes of illustrators. Our work seeks to bridge previous research conducted with non-art games to affect psychological outcomes, particularly literature on how games can foster empathy~\cite{isbister2016games, kaufman2012changing}, emotional regulation~\cite{villani2018videogames}, and social learning~\cite{isbister2016connecting}. Our work builds on these themes, contributing game mechanics that address the psychological impacts of self-criticism in the context of illustration. Additionally, while prior research has advanced our understanding of designing games for diverse communities and their specific needs~\cite{hantsbarger2022alienated, to2018character}, we focus on how to effectively design games for artists and illustrators and explore game mechanics that support them in managing self-evaluation. 

\section{Art Card Game}
\label{sec:system}

\begin{figure*}[tb]
  \centering
  \includegraphics[width=0.78\textwidth]{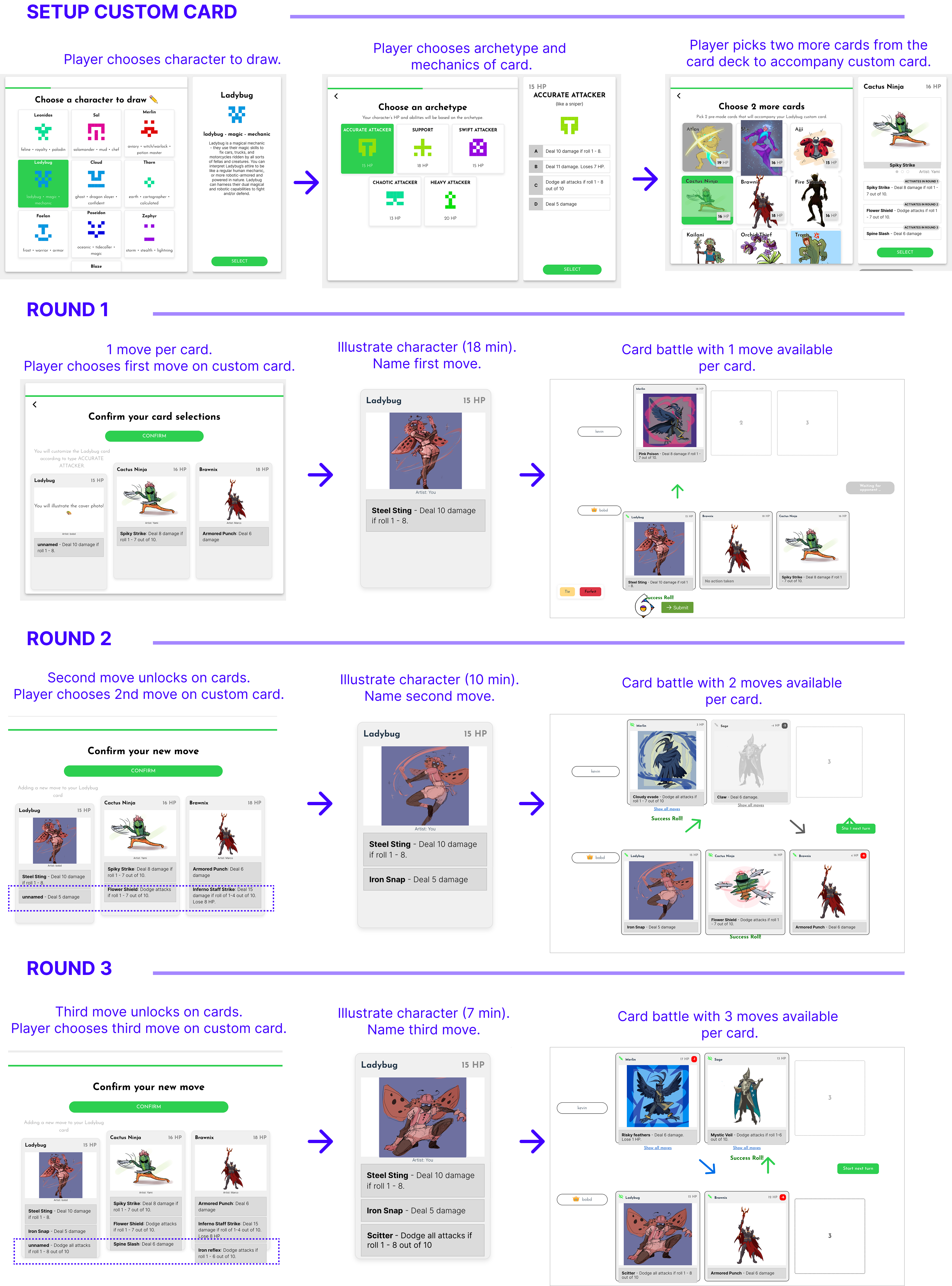}
  \caption{Player journey in ACG. In phase one, the player setups their custom card, which includes picking the character description and card archetype (card's HP and moves). In this phase, the player also selects two cards from the card deck to accompany the custom card. After the card setup phase, the player plays the card game against their opponent in three rounds. During each round, the player must select a new move for the custom card and illustrate the character depicting that move selected. After illustration and move selection, the card game begins.}
  \Description{A sequence of screens describing the player's journey in Art Card Game. The figure is arranged in four horizontal sections (rows): Setup Custom Card, Round 1, Round 2, and Round 3. The Setup Custom Card row contains three interface screenshots showing the steps for constructing a custom card. Across all three, the interface is divided into a larger left panel containing a grid of options and a narrower right panel showing a preview with details of the currently selected option. In step 1, there is a grid of ten character templates, each with a name, small icon, and descriptive keywords. The right panel shows the selected character with a short textual description. Step 2 presents five archetypes, each with an archetype name, icon, and hit-point value. The right panel displays the selected archetype and a list of four potential move effects. Step 3 shows eight premade cards, each showing an illustration, character name, and hit-point value. The right panel shows the selected card’s illustration and the three moves associated with that card, each labeled with its name, effect, and activation round. The next three rows represent Round 1, Round 2, and Round 3. Each row contains three screenshots corresponding to the same sequence of actions: (1) selecting a new move for the custom card (with accompanying cards also unlocking a new move), (2) illustrating the custom card to match that move, and (3) entering a card battle in which each card can use the moves unlocked by that round. In Round 1, the move-selection screen shows the custom card and two premade cards, each with a single move. The illustration screen displays the custom card with its first move and artwork. The battle screen shows all three cards with one available move per card. In Round 2, the move-selection screen again lists the three cards, now each with two moves. The illustration screen shows the updated custom card with new artwork reflecting its second move. The battle screen depicts gameplay with two moves available per card. In Round 3, the move-selection screen lists the same three cards, now each with three moves. The illustration screen presents the final custom-card artwork corresponding to the third move. The battle screen shows all three cards with full access to their three moves.}
  \label{fig:game-flow}
\end{figure*}

Art Card Game (ACG) is a synchronous, competitive two-player online trading card game that embeds illustration \emph{inside} a consequential play loop. Each session, players select a character prompt and an attack style, choose a move, and draw a cover image under a fixed time limit depicting that move; the illustration becomes the card’s cover and is immediately used in a head-to-head battle (Figure \ref{fig:game-flow}). \newcc{Through evaluative off-centering,} ACG reduces the emphasis on aesthetic judgment during creation and helps lessen momentary self-criticism.

\subsection{Design of Art Card Game}
\newcc{To instantiate evaluative off-centering, we embedded illustration within an intrinsically motivating surrounding activity---a competitive card game---with three elements: social interaction, art-as-ticket (illustration enables participation rather than determines success), and strategic play (competitive decision-making aimed at winning). Illustration is integrated across these elements to keep WIPs oriented toward participation and play rather than aesthetic evaluation.}

\subsubsection{\newcc{Designing an intrinsically motivating surrounding activity}}
\newcc{We began with two commitments: off-centering would likely benefit from social interaction, since solitary making can leave room for evaluative rumination, and integrating illustration-making into the surrounding activity. However, the concrete form was underspecified. Although prior work suggests social support can make creative effort feel more shared and sustainable~\cite{nea2024arts,carron1996social,lin2024socialsupport,neff2023review}, we did not yet know which social structure (e.g., collaborative or competitive) would best off-center evaluation, nor which surrounding activity could reliably hold attention.}

\newcc{We prototyped narrative and collaborative formats (RPG-like quests, telephone comics, shared canvases) with small playtests (2--4 artists). \new{Across these prototypes, we preserved artists’ preferred drawing tools, stylistic preferences, and choice of subject matter while varying the brief, timing, social setting, and output format (e.g., a comic). The RPG-like quest and shared-canvas prototypes allowed artists to draw until they felt finished; the telephone-comic prototype introduced a timer as the primary constraint.} However, narrative immersion raised the stakes of ``fitting'' the story, leading to requests for more time to polish. Collaborative formats repeatedly re-centered evaluation and introduced logistical frictions. Shared artifacts diluted ownership and style/finish mismatches increased harsh judgment and disposability. The coordination overhead (waiting, turn dependencies) stalled momentum, creating gaps where self-critique resurfaced.}

\subsubsection{\newcc{Why a competitive card game}}
\newcc{A key turning point was shifting to strategic, competitive play, which reduced dependency (parallel work), preserved ownership, kept time constraints intact (``fighting the clock'' matched the competitive spirit), and sustained engagement through head-to-head battling.}

\newcc{The card game direction emerged from a simple design prompt: \emph{what already inspires illustrators to create character art?} We looked to trading card games, which often depict character-focused illustrations as card art. We distilled familiar TCG mechanics~\cite{chen2009} to keep gameplay learnable in a single sitting so strategy would not overshadow illustration.}

\newcc{Cover art naturally links to what a card represents and does. In ACG, we translated this idea into card customization. Players build a card and for every chosen card move, they illustrate a character pose depicting the chosen move. In this way, illustration is reframed as a ticket necessary to build a usable game piece rather than producing polished artwork, helping keep evaluation off-centered.}

\subsubsection{\newcc{ACG core elements}}
\newcc{Iterative prototyping converged on the ACG format that keeps the surrounding activity intrinsically motivating and prevents evaluation from drifting back onto art quality. Prototyping clarified the social interaction: synchronous (shared session), independent (each person retains ownership), and competitive (head-to-head without co-authoring a single artifact, which previously amplified style-mismatch and ``quality'' pressure). Within that structure, art-as-ticket makes illustration a required input: each round, players create cover art for a customizable card that depicts the selected move, producing a usable, expressive game piece rather than a polished standalone image. We set the ticket’s quality bar at ``good enough'' with time limits and reference art via a pre-made deck that demonstrates quality. Finally, ACG embeds these tickets inside strategic play grounded in simplified trading card mechanics that remain engaging while keeping evaluative attention on gameplay rather than illustration.}

\subsection{Core Loop: Game Mechanics and Play Dynamics}
ACG is a synchronous, two-player competitive game played over three rounds. Each round has two phases: (1) \emph{card customization} (choose a move and illustrate it under a time limit) and (2) \emph{head-to-head battle} (play cards and resolve moves). Players win a round by reducing all three of the opponent’s cards to 0 HP; the overall winner is the player who wins the most rounds.

\paragraph{Pre-round setup (5--10 min).}
Before Round 1, each player selects a hand of three cards used throughout the session: one \emph{custom card} and two \emph{pre-made} cards from the shared deck (Figure~\ref{fig:game-flow}). To configure the custom card, players choose (i) a character prompt, (ii) an attack style (\emph{archetype}) that defines HP and a set of moves, and (iii) the first move (used in Round 1).

\paragraph{Rounds 1--3: customize, then battle.}
At the start of each round, the custom card gains a new move. Players select that move from their archetype’s move set, then illustrate a cover image that depicts the character performing that move. Illustration time decreases across rounds (18 min, then 10 min, then 7 min); when time expires (or earlier), players upload their illustration image, which becomes the move’s cover art on the custom card.

After both players upload, the battle phase begins. Players take turns selecting up to one move per card currently in play and choosing targets on the opponent’s side. Many moves are deterministic, while others depend on a 10-sided die roll (e.g., dodge/reflect effects). Each turn includes a private planning step followed by a reveal/resolution step where both players’ choices are shown and executed. Players may declare a tie in a stalemate or forfeit if the outcome is inevitable.

\section{Evaluation: Method}
To investigate whether \emph{evaluative off-centering} effectively mitigates momentary self-criticism compared to solitary illustration, we conducted a four-day, between-subjects, randomized controlled trial. We measured participants' self-reported pride, enjoyment, and excitement.

\subsection{Participants and Attrition}
Over a week, we recruited active hobbyist illustrators (18+) via university lists and Discord art communities, foregoing demographic collection to preserve trust~\cite{discord-doxxing-policy}. Of 97 screened candidates, 51 confirmed and were randomly assigned to ACG or control conditions. On day 1-2 of the study, seven and ten participants dropped out of the ACG and control, respectively. Since partnership was crucial to ACG, we utilized waitlisted illustrators and hired three on-call Upworkers to fill vacancies (Upworker submissions excluded from data analysis). We briefly attempted re-recruitment, adding one control participant, but ultimately ceased efforts to prioritize managing ACG's complex scheduling and technical stability. Ultimately, 38 participants (23 ACG, 15 Control) completed the four-day protocol.

\subsection{Measures}
After each daily session, participants completed a daily survey to upload their artwork and rate three measures (1--7 Likert): \textit{Enjoyment} (``I enjoyed today’s activity''), \textit{Pride} (``I feel proud of the art I produced today''), and \textit{Excitement} (``I’m excited for tomorrow’s activity''), each with a brief open-text explanation. Pride, enjoyment, and excitement are measures adapted from the PANAS scale~\cite{watson1988development}, which are inversely associated with self-criticism~\cite{zuroff2016conceptualizing}. \new{We selected these three affective measures rather than administering the full 20-item PANAS scale to reduce participant burden in the four-day repeated-measures study and to focus on outcomes most directly tied to our research question around self-criticism. The full PANAS scale captures broader positive and negative affective states, including items outside the scope of our study such as measuring feeling ``guilty'' and ``afraid''.}

\subsection{Procedure}
Participants completed one session daily for four days. ACG participants met a scheduled partner over Discord/Zoom for $\sim$60 minutes to play the card game, where the illustration activity was included. The control group functioned as an active control, performing the exact same illustration tasks, but without treatment (no partner nor gameplay). To equalize session length to 60 minutes (while matching the art-theme without introducing skill-based evaluation), control participants watched 25 minutes of art-historical documentary videos after drawing. \new{We selected these videos to maintain thematic continuity with artmaking, rather than introducing non-art-related filler content. Given that many participants were digital artists, we avoided popular digital art content (e.g., tutorials, speedpaints, livestream illustration, art motivational talks, or critique sessions) that could invite skill comparison. The videos provided a broad survey---from prehistoric artifacts and ancient sculpture to 20th-century modernism and abstraction---selected to be chronologically and stylistically distinct from participants’ digital character-illustration tasks.}

An end-of-study survey collected qualitative feedback. Additionally, participating in an optional 30-minute voice-only interview counted towards compensation. Participants received up to a \$76 gift card (prorated by completion).

\begin{figure*}[tb]
  \centering
  \includegraphics[width=.75\linewidth]{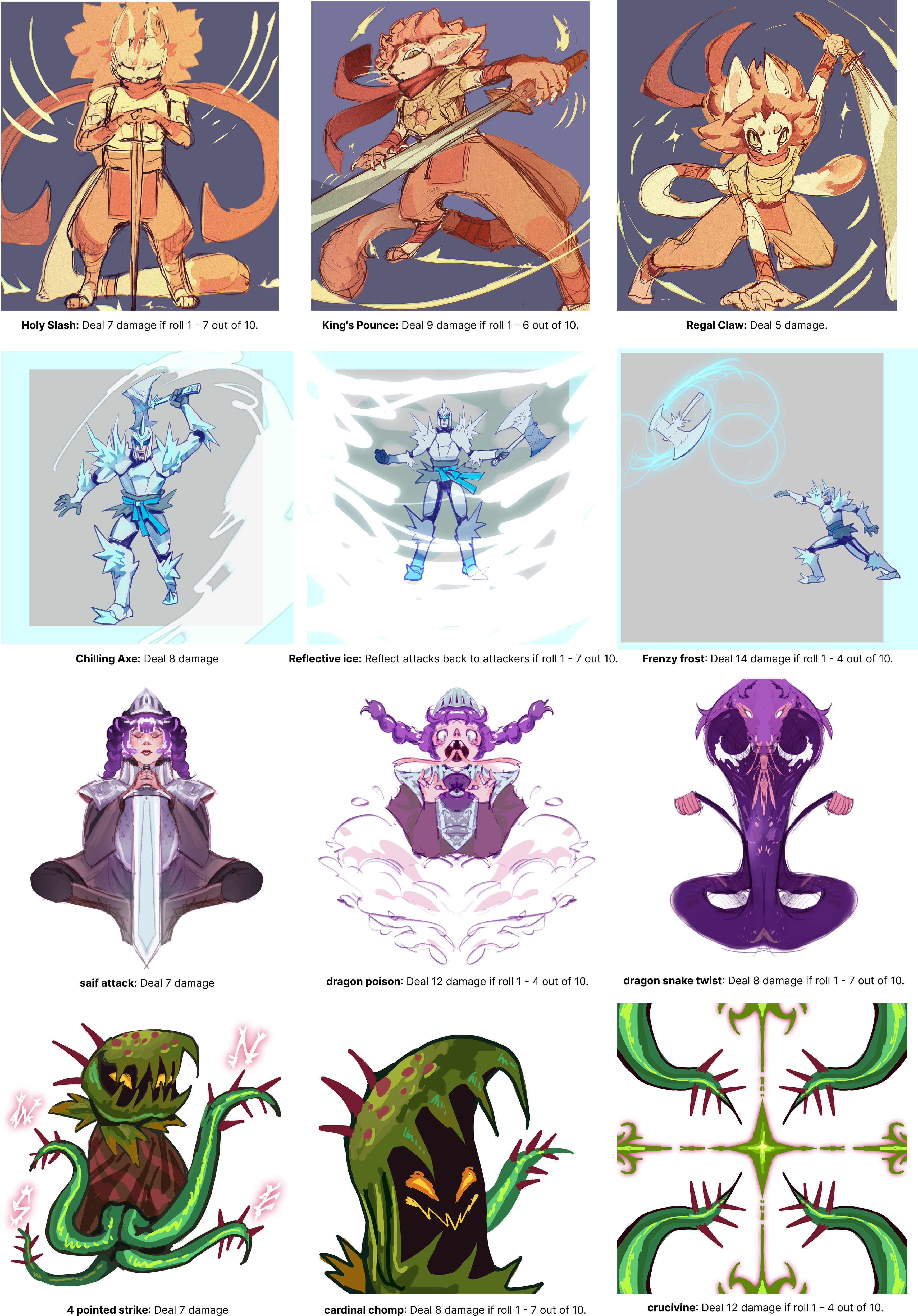}
  \caption{Examples of participant artwork from ACG. Each row (a different artist) shows the same character redrawn across three rounds in one game session. The character pose in each image depicts the move used that round, including the artist-named move title (e.g., ‘Chilling Axe’) and its mechanic (e.g., deal 8 damage). Images are ordered by the decreasing time limits from left to right: 18 minutes (round 1), 10 minutes (round 2), and 7 minutes (round 3).}
  \Description{Examples of participant artwork across three rounds, showing how each character is redrawn under decreasing time limits. The figure is a grid of four rows, with three illustrations in each. The three illustrations correspond to an artist's rendition of the same character redrawn across three rounds in one Art Card Game session. Row 1 showcases a feline warrior. In the first image, the warrior is standing still and holding a downward-pointing sword; the second shows the warrior in a wide stance wielding their sword across their body; the third shows the warrior crouched down with their sword raised. Row 2 exhibits a spiky, armored ice-themed soldier: the first image shows the soldier walking forward holding their ice axe overhead; the second image positions the soldier in a power-up stance surrounded by swirling energy; the third shows them throwing their axe across the scene. Row 3 contains a crowned, magical knight. The first image shows the knight seated in a calm pose holding their sword faced down; the second shows them exhaling a poisonous mist from their mouth; the third shows the knight transformed into a serpent-like dragon. Finally, row 4 posses a spiky plant monster with vine-like limbs. In the first image, the monster is posing with its four vine-like limbs; the second presents the monster about to chomp with its sharp-toothed mouth; the third shows a radial, mirrored arrangement of its vine-like limbs.}
  \label{fig:all-artwork}
\end{figure*}

\begin{figure*}[tb]
  \centering
  \includegraphics[width=.6\linewidth]{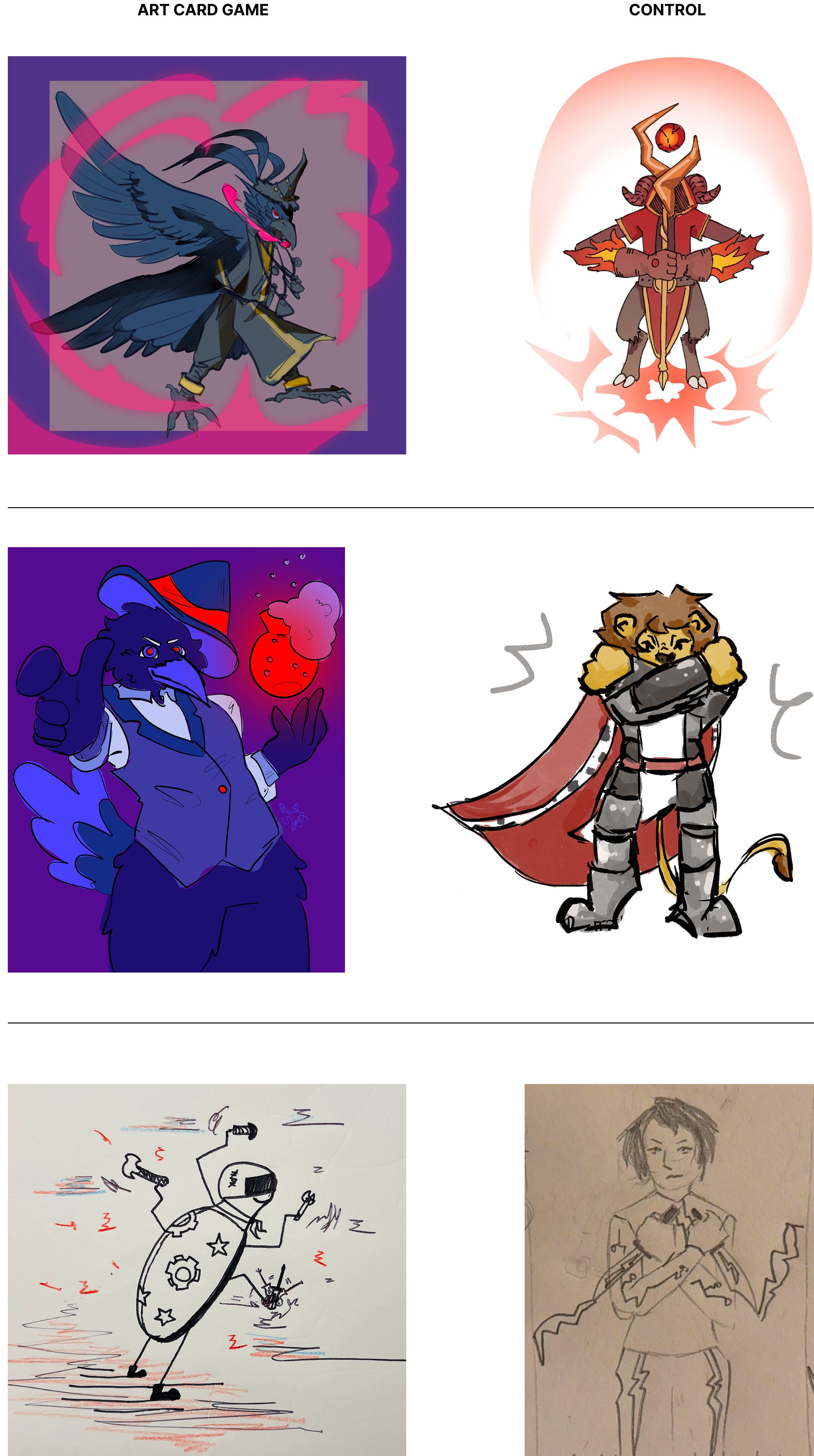}
  \caption{Examples of participant artwork from Art Card Game (left column) and the control (right column) conditions. Each illustration was produced by different artists under the 18-minute time constraint. Rows show three types of outcomes that emerged. The top row contains colored and highly detailed pieces, the middle row shows colored pieces with simpler detail, and the bottom row shows primarily line-based or minimally colored sketches.}
  \Description{Examples of the quality and styles of participant artwork from Art Card Game and the control conditions. The figure contains three rows with two illustrations in each; the left illustration belongs to the ACG condition and the right to the control. The top row contains two illustrations that are colored and highly detailed. The top row’s left image contains a crow-like magician surrounded with smoke and facing from one side with an outstretched wing. The right image shows a fiery, bighorn sheep sorcerer standing center, holding a large staff topped with a red gem; flame-like shapes radiate from the sorcerer. The middle row contains two illustrations that are colored with simpler details. The middle row’s left image displays an anthropomorphic crow wearing a vest and hat, with one hand pointing outward while a glowing red potion levitates in the other hand. The right image shows a lion-like knight dressed in armor with a long red cape, standing with arms crossed. The bottom row contains two illustrations that are line-based and minimally colored. The left image is a paper and pen line-drawing of a ladybug mechanic. The ladybug’s body is decorated with stars and gears, and the ladybug is waving its limbs that are holding different home improvement tools. The right image is a pencil sketch of a human character crossing their arms across their chest; each hand clasps a knife of electrical energy.}
  \label{fig:compare-acg-control}
\end{figure*}


\subsection{Data Collection}
At the end of the study, we had 146 daily survey submissions across both conditions, 438 art submissions, and interview notes from 19 of 23 participants in the ACG condition who consented to an audio interview. For the daily survey submission, each survey response contains daily Likert ratings (1–7) of Pride, Enjoyment, and Excitement along with short-form text response justifications. We use this information to supply our quantitative analysis of our findings. Moreover, we complemented our quantitative findings with a qualitative analysis. We analyzed participants' (i) daily open-text reflections attached to the Pride/Enjoyment/Excitement Likert ratings and (ii) post-study interviews with participants in the ACG condition.

\section{Results}

Thirty-eight participants completed the study, yielding 146 daily survey submissions. Most participants completed all four days (Control $n=12$, ACG $n=20$), while a smaller number were only able to complete three days (Control $n=2$, ACG $n=3$) due to scheduling issues. We performed follow-up semi-structured interviews with nineteen of the participants.

A sample of the artwork produced in ACG and the control conditions can be found in Fig. \ref{fig:all-artwork} and Fig. \ref{fig:compare-acg-control}, respectively.

\paragraph{Modeling approach.} We fit cumulative link mixed models (ordered logistic) for Enjoyment, Pride, and Excitement Likert ratings, with fixed effects for Condition (ACG vs. Control), Study Day (1–4), and an indicator variable for the presence of a game-breaking bug during the session (nested in condition), plus a random intercept for participant (146 observations from 38 participants). This approach (i) accounts for within-participant dependence across days, (ii) handles unbalanced repeated measures (e.g., three vs. four days completed), and (iii) allows fixed effects for day and a session-level bug indicator. We report odds ratios (OR), 95\% CIs, and p-values; OR > 1 indicates higher odds of a higher rating in ACG vs. control. Descriptive distributions are shown in Fig.~\ref{fig:likert-overall}; model estimates appear in Fig.\ref{fig:forest} and Table~\ref{tab:clmm_three_stars}. 
In the ACG condition, ACG website bugs occurred intermittently at the session level—e.g., card-selection errors and stalled game states—such that some participants encountered them on some days, while others did not. \footnote{The issues were persistent enough to warrant a ~22-hour pause of the ACG condition and implement fixes on the second day of the study.} To quantify the extent to which technical disruptions influenced participants’ reported experience, we manually coded each daily submission for the presence of the bug. Coding was based on participants’ free-text responses, which often explicitly mentioned technical problems. Each submission was thus annotated with a binary indicator (1 = bug reported, 0 = no bug), which we included in the statistical model. About 28\% of ACG sessions were coded with the presence of a game-breaking bug.

\paragraph{Quantitative Results Overview.} ACG produced reliably higher ratings than the solo control on Enjoyment ($\beta=2.215$, $\mathrm{OR}=9.16$, $p<.001$) and Pride ($\beta=1.693$, $\mathrm{OR}=5.44$, $p<.01$). For Excitement, the condition effect was positive but not significant ($\beta=1.162$, $\mathrm{OR}=3.20$, \new{$p=.116$}) at our level of statistical power. These results are visualized in  the forest plot (Fig.~\ref{fig:forest}) and reported in Table~\ref{tab:clmm_three_stars}.

\input{likert-by-condition}

\input{forest-plot}

\input{clmm_three_models_with_day_stars_from_summary}

\subsection{ACG increased feelings of pride towards one's artwork}
As described above, ACG participants reported greater pride toward their artwork (Fig.~\ref{fig:forest}; Table~\ref{tab:clmm_three_stars}). In practical terms, participants in ACG had over five times the cumulative odds of reporting feeling prouder of their day’s artwork than those in the solo control condition.

Technical issues did not undermine the pride result. Sessions with a reported bug showed lower odds of a higher pride rating, but the effect was not statistically significant.

\subsubsection{Qualitative context.}
Open-ended responses and interviews echoed the quantitative effect: ACG participants more often described pride in their daily illustrations, using language of ownership towards art, celebrating progress and improvement, consistently meeting ones' artistic expectations, and experimenting with new styles (and succeeding). Below are the recurring themes, with representative examples.

\paragraph{Stronger sense of ownership and affection for artwork} 
Players expressed a sense of ownership towards their illustration--sometimes talking about their cards and characters as \emph{theirs}, rather than as disposable outputs of a timed task. Comments indicated such signs of affection or possession: “It’s so cuteeee, I love Merlinnnn,” “I really loved the character I created today,” and “I love my little guy.” By contrast, control participants more often framed pieces as roughs or work-in-progress (“I don’t take pride in rough drawings/paintings”; “These are just rough ideas in digital form”) and used more self-critical language ("I think that the character looks cute enough to cover up the fact that I had no clue to what I was doing"), with milder signs of approval or affection (“I think the character I drew was cute, so I’m satisfied with that”) in comparison to those in ACG.

\paragraph{Willingness to experiment—and satisfaction when it worked}
Pride gains appeared through ACG participants' indications of artistic experimentation, especially when successfully executed in the illustrator's opinion. Many illustrators arrived in the study knowing how to execute a particular art style, but chose to confidently adapt and experiment in response to the ACG task. One participant viewed it as "chang[ing] my art style to be more efficient" while still feeling it was "really fun." This particular participant made numerous changes from their normal workflow such \new{as} using "three brushes" to "one brush", "not needing any reference references at all," and "just freestyled everything". Other participants exhibited pride through experimentation by tackling different artistic subject matter they typically avoid: "glad i was able to get more out of my comfort zone by drawing some sort of variation of a mech/armored outfit as I stray away from harder textiles." 

\paragraph{Less pride in control condition tone} Control participants were not uniformly negative about their illustrations, but their comments leaned more toward focusing on illustration incompleteness and judgment by viewers. Control participants tended to view their results like incomplete works-in-progress rather than a progress gain or sign of ownership ("I don't feel proud, my art looks like children's drawing" or "why did this take 7 minutes? This looks like a 3 minute doodle"). Moreover, the control condition presented a few instances of self-consciousness and concerns of external judgment, which was not present in the ACG condition at all. Examples include: "It is not something I would show to others as it does not meet my standards, its rushed, without clear direction or design" and "I feel embarrassed in how bad some of them are, and its not something I would want to show anyone." This contrasts with ACG participants’ frequent mentions of meeting internal quality standards and even willingness to experiment with different artistic styles.
 
\subsection{Enjoyment was higher when drawing was embedded in gameplay than in solo illustration}
Participants enjoyed ACG more than solo illustration. 
The statistical model corresponded to roughly nine-fold higher cumulative odds of a higher enjoyment category under ACG. Unsurprisingly, reported technical issues sharply attenuated enjoyment within ACG: bug-flagged sessions had about one-fifth the odds of a higher enjoyment rating relative to non-bug sessions (\emph{OR} \(\approx\) 0.20, 95\% CI below 1, \(p<.001\)). The condition advantage remains positive overall but is masked in sessions affected by game-breaking bugs.

\subsubsection{Qualitative context.} 

Enjoyment in ACG arose from the card game battles, partner interaction, illustration task, or some of combination the three, and the weighting of these enjoyment sources varied by participant. We identified three recurring enjoyment profiles: (1) \emph{gameplay-focused participants}, who frequently enjoyed everything about ACG---in particular, the head-to-head card battle---and often reported high enjoyment even on bug-affected days; (2) \emph{partner-focused participants}, whose enjoyment rose with conversational, engaged partners; and (3) \emph{low-social participants}, who derived satisfaction chiefly from the illustration tasks and playing the card game. In most cases, ACG sessions with technical issues affected participant's enjoyment.

\emph{Gameplay-focused participants:} In the end-of-study survey, five participants reported playing the card game as their most enjoyable activity in the ACG and appeared unphased by the presence of bugs. One player expressed during their interview, "I loved playing that game with all my heart \ldots I loved everything about it. I wasn't bored with it like at all \ldots I would want to play more." Gameplay-focused participants consistently expressed strong desire to continue to play the game after the study was over (e.g. "Will there be a way that I can play it again with maybe one of my another friend?," asked a gameplay-focused participant) or expressed that they would recommend their friends to play it.

(2) \emph{Partner-focused participants.}
Eleven participants reported meeting a new artist as their most enjoyable activity in the ACG and highlighted the social interactions, whether on-topic to ACG or off-topic personal discussions, as central to their enjoyment. These participants sometimes reported enjoying the card game play, especially playing with their partner, or they strongly disliked the card game due to adverse experiences with the games' bugs, disliking the game mechanics, or both. Everyone still enjoyed the illustration task.  

Players who liked both their partner interactions and the card game emphasized the ``call with my partner while playing the game'' made them feel ``more connected with the person.'' They ``didn’t really care about winning or losing'' the game and really ``just enjoy[ed] talking with the person.'' A few participant pairs continued to speak with one another after their daily session---for one to two hours in some cases---and exchanged contact information. 
Moreover, some of these player profiles disliked partners who seemed disinterested in social engagement and whose conversations focused on the logistics of completing the game. 

(3) \emph{Low-social participants.} Low-social participants were participants who reported that they did not engage too much socially with their partner and therefore did not derive enjoyment from their partner, but instead from \new{the} illustration and card game. Low-social participants preferred not to socialize due to constraints (e.g. losing their voice because of sickness or staying muted to block out external noise), preference for silence, or reservations to connect socially. One player described some of their partners as ``more reserved I guess, like me. I think we were all a little embarrassed to talk.'' 

\subsection{ACG gameplay effectively decentered self-criticism during illustration}
In interviews and daily texts, participants in the ACG condition consistently framed drawing as making something to use immediately in play, rather than a standalone sketch to judge for themselves. The two themes below illustrate this decentering: drawings became personal game assets and illustration choices were anchored to play goals.

\subsubsection{Drawings were treated as personal, playable assets (not art pieces)}
Relative to the control, participants in ACG framed illustration not as an isolated sketching task but as building a playable asset (a customized character avatar) for the upcoming card battle. Many described this immediate in-game use as especially fun and motivating: ``I think there is joy in using the character you just drew in a game," said one. ``You customize this character \ldots like an avatar and you just like play it in the game. It feels a lot more fun than \ldots if you ha[d] a stock character." Another called this the standout moment of the study: ``The coolest, dopest part of the game was that you can draw your own character. \ldots It feels like a movie or a video game \ldots because it's like boom, you just created a character and boom, now it’s on screen \ldots that was a really good experience. Never experienced that before.” Together, these reports indicate that participants viewed their drawings as game pieces they would immediately use, not as standalone practice sketches.

\subsubsection{Art decisions were tied to play goals, redirecting attention from polishing illustration to gameplay use}
In ACG, participants made illustration decisions in terms of imminent, social and strategic use in the game (how it will look and perform on the battlefield), rather than as means to illustrate quality pictures like in the control condition. One player described weighing “whether to pick a move because it’s actually really strong or … because I haven’t drawn it before,” typically opting to prioritize game strategy over illustration and therefore chose ``[moves] that I knew [were] strong." Another participant took the opposite approach, saying winning the battle ``didn't matter," so they ``prioritized ... picking an interesting character to draw and ... moves that I thought were not useless but also might be interesting to draw." Consequently, during card battling with their partner, they ``would try to play my new move just to show off the art" to their partner. 

\subsection{Drawbacks of Art Card Game}

\subsubsection{Mechanical depth and balance}
Some participants---especially those familiar with complex TCGs---found the battle system too simple. They also raised balance concerns: a few moves felt overly dominant due to generous dice windows, while other cards and moves felt comparatively weak or ``useless.'' Several participants noted that later-unlocked moves did not always feel stronger or more distinctive, and that limited move types (attack/dodge/reflect) made archetypes feel less differentiated and reduced opportunities for synergies across cards. Reactions to randomness, via dice rolls, were mixed: some enjoyed dice as playful uncertainty, while others felt outcomes depended too heavily on chance, making rounds less satisfying.

\subsubsection{Pacing and Social Dependence}
Beyond balance and depth, some participants found the decreasing drawing timers stressful early on, and several disliked idle time during battles. Enjoyment also varied with partner interaction: some highly valued conversation, while others preferred minimal social interaction.


\section{Discussion}
\newcc{Through our evaluation of ACG, we return to a central design question: how can we embed creative production in a surrounding activity so that works-in-progress (WIPs) are less likely to trigger evaluation and self-criticism? Our findings suggest the goal is not simply to ``distract'' illustrators from judgment, but to make the production process feel intrinsically worthwhile and fun even when the result is still rough. Across our findings, off-centering showed up in two related outcomes: participants felt better about (1) the resulting WIPs (e.g., pride and ownership toward rough artifacts) and/or (2) the production process itself (e.g., confidence and willingness to experiment).}

\newcc{In our study, ownership and pride toward WIPs served as a useful indicator of effective evaluative off-centering: when participants feel good about an in-progress artifact, they are more likely to finish it, iterate on it, or treat it as a meaningful attempt rather than a disposable step toward a polished outcome.}

\newcc{In this section, we reflect on how three design elements---social interaction, art-as-a-ticket to entry, and strategic play---made works-in-progress feel less evaluative and more rewarding to produce, both by supporting experimentation and pride in the resulting artifacts and by sustaining engagement through the surrounding activity (e.g., enjoyment of social play and/or mastering ACG gameplay). Importantly, evaluative off-centering depends on what elements participants find intrinsically motivating in the surrounding activity.}

\subsection{\newcc{Reflections of evaluative off-centering}}
\newcc{We structure the following reflections around three elements of ACG’s surrounding activity---social interaction, art-as-a-ticket to entry, and strategic play---and discuss how each affected illustrators and redirected self-evaluation. While we attempt to isolate these elements in terms of discussion, note that they often worked in conjunction with each other; we argue that it is the combination of all three that contributes to the observed effects.}

\subsubsection{\newcc{Social Interaction for Evaluative Off-Centering}}
\newcc{In our study, social interaction did not operate as a single off-centering mechanism; instead, we observed a bundle of social pathways---art reveals, artistic style exposure, conversation---that can each sustain WIP-making and off-center self-criticism in different ways.}

\paragraph{\newcc{Art Reveals.}} \newcc{Embedding the art reveal in gameplay, rather than presenting it as a standalone showcase, positioned the WIP as something to react to, celebrate, and immediately use in ongoing play rather than something to pause and evaluate for artistic quality. After each round’s illustration phase, once the card game began, players encountered a partner’s illustration as the illustrated cards were played. This created a brief celebratory moment wherein partners exchanged quick compliments and admiration before continuing playing. We argue that embedding the reveal into the surrounding activity gives partners, or even a community, an opportunity to acknowledge and reward WIPs while naturally redirecting attention back to the shared activity.}

\paragraph{\newcc{Style Exposure.}} 
\newcc{Unlike the social reveal, which often cultivated pride in the (unfinished) outcome, style exposure---noticing the stylistic choices in a partner’s artwork---shaped how participants felt about the production process, increasing confidence to experiment and motivation to keep practicing. Because partners drew from the same pool of prompts, encountering a partner’s artwork often highlighted different directions one could take, prompting reflection on one’s own choices and increasing confidence to experiment. One participant's comment captured this pattern: ``it encouraged me ... to create creative ways of making because I hav[e] a very strict card design ... like Pokémon sort of cards ... but seeing how [my partner] gave their own twist to it [painterly style] inspired me.'' This participant later described trying their own version of the painterly style and feeling satisfied with the result.}

\newcc{For some participants, style exposure created social competition or comparisons that have a more positive outcome: a self-improvement stance that motivated continued practice. One participant practiced drawing outside the ACG sessions to keep up with their partners’ skills and feel more prepared to perform well. Nevertheless, others reported being largely unaffected by their artwork being seen, relying instead on their own internal standards when evaluating their art. We interpret this observation through prior distinctions between self-critical \emph{self-correction} (aimed at improvement) and \emph{self-persecution} (aimed at harming the self)~\cite{gilbert2004criticizing}. Depending on the system of experience design, style exposure could prompt self-corrective motivation without necessarily reintroducing self-persecuting comparison. For example, when artifacts are primarily presented for viewing in showcase- or curation-oriented settings---whether featuring polished work (e.g., Instagram) or works-in-progress---posting can begin to feel like public performance and create anticipation of judgment. In contrast, by designing for evaluative off-centering, a surrounding activity reframes creative artifacts as part of ongoing interaction rather than standalone posts awaiting evaluation.}

\paragraph{\newcc{Conversation.}} \newcc{Conversation---often occurring around rather than during timed illustration---helped maintain momentum, reducing opportunities to dwell on self-critique and keeping attention oriented toward the shared experience and what came next. Because the illustration phases were time-bounded and attention-demand- ing, many participants described socializing primarily between rounds, during transitions, or while reacting to play, which made the overall session feel more enjoyable and socially energizing. Importantly, conversation was not equally central across participants: some chatted minimally (or not at all) often due to shyness or low desire to socialize. Though these less social participants still appeared to off-center evaluation through other pathways (e.g., focusing on gameplay mastery or the artifact-centered reveal/style dynamics). This variation suggests value in adjustable social intensity, such as text-first interaction, lightweight reactions, or lower-social, lurker participation.}

\subsubsection{\newcc{Art-as-a-ticket for Evaluative Off-Centering}}
\newcc{Art-as-a-ticket design---where players must create an illustration in order to par- ticipate---can shape participants’ attitudes toward their WIPs by reframing them from aesthetic artifacts to judge into (1) a requirement for entry and (2) purposeful artifacts that matter within a surrounding activity (here, gameplay). Crucially, treating art as a ticket differs from treating art quality as the metric of success: when winning and losing are determined by the surrounding activity rather than by aesthetics, participants have fewer reasons to fixate on polish and can instead pursue the activity’s goals. Further, in our design the ticket did not end at entry. Because illustrated cards were immediately \emph{used} in play, participants described sketches ``came to life'' as usable game pieces, appearing on the card battlefield, persisting across rounds as personal assets, and shaping decisions during the match. At a high level, giving WIPs an immediate purpose helped reduce feelings of meaninglessness or disposability that can accompany rough drafts. This contrasts with party drawing games (e.g., Gartic Phone), where sketches may be intentionally disposable due to a low-fidelity, joke-oriented frame and extremely short time limits (e.g. matter of seconds).}

\newcc{Overall, these evaluative off-centering benefits depend on how the ticket is designed. If the implicit expectation becomes polished artwork, the requirement can instead intensify evaluation. In our design, we established a ``good enough'' quality bar for the ticket by (1) exposing players to a pre-made card deck that established aesthetic expectations and (2) using timers to discourage extended polishing while still leaving room for personal expression.}

\subsubsection{\newcc{Strategic Play for Evaluative Off-Centering}}
\newcc{Strategic play---making competitive decisions aimed at winning within the game---off-centered evaluation by shifting attention toward mastering the game, which eased both in-process and outcome-focused judgment of their WIPs. For some participants---especially those less motivated by social connection---the desire to play well and improve at the card game functioned as a productive distraction: it absorbed attention during and between rounds and offered an alternative basis for competence that did not depend on illustration quality. In this sense, mastery motivation can reduce the salience of self-critique by giving participants something else to optimize for, and in some cases participants explicitly preferred focusing on gameplay over the illustration itself.}

\newcc{One possible explanation for this game-first orientation is our intentionally loose coupling between illustration and game logic. Because illustration quality did not affect who won, some participants could treat drawing as a lightweight prerequisite and devote attention to gameplay. At the same time, our early prototypes suggest a tradeoff: when the content or perceived quality of illustrations affected power-ups or win conditions, the tighter coupling centered attention on the illustrations and reintroduced evaluative pressure---and sometimes dispute---between partners. More generally, the tightness of coupling between illustration and the surrounding activity is a design parameter: a stronger coupling makes illustration more directly affect the activity outcomes, but may also increase the risk of re-centering evaluation.}

\newcc{Even under the same coupling design, participants related illustration and gameplay decisions in markedly different ways. Some adopted a strategy-first orientation, prioritizing strong moves even when they were less interesting to draw, while others took a drawability-first orientation, selecting moves that afforded richer artistic expression even if they were not optimal strategically. This divergence suggests a key design choice: systems can be tuned to privilege gameplay mastery, artistic exploration, or a balance of the two. In our case, the limited move set (e.g., attack, dodge, reflect) may have accentuated this tradeoff by constraining the space of ``strong but also fun-to-draw'' options; expanding and diversifying the move set could better support players who want to play competitively while still having satisfying, varied illustration opportunities.}

\subsection{Broader Implications}
\newcc{Taken together, these elements suggest evaluative off-centering as a design space for creativity support: surrounding activities can reduce self-criticism by helping people feel better about (1) their WIP outcomes (e.g., pride and ownership toward rough artifacts) and/or (2) the production process itself (e.g., confidence and willingness to experiment). Designers can support these outcomes through a bundle of surrounding-activity levers: for example, (i) creating brief, celebratory moments around WIPs without turning them into showcases, (ii) reframing rough artifacts as meaningful in use rather than judged in isolation, (iii) providing time-bounded momentum that limits polishing windows, and/or (iv) offer adjustable social pathways that reward participation while accommodating low-social preferences. Rather than relying on any single feature (e.g., ``add social''), ACG illustrates how combining multiple off-centering elements can support user-centric benefits~\cite{beyondproductivity} while preserving artists’ agency over what and how they draw.}

\subsection{Limitations}
\subsubsection{Boundary Conditions}
Evaluative off-centering depends on whether the surrounding activity is intrinsically motivating for a given creator. Social interaction and voice conversation can be engaging for some but uncomfortable for others, suggesting value in lower-social modes (e.g., text chat, lightweight reactions). Likewise, competitive game formats will not suit everyone. 

\new{Not all game mechanics produce evaluative off-centering: mechanics that score, rank, or compare artifacts by aesthetic quality---for example, leaderboards or voting systems for the ``best'' drawing---would re-center evaluation on the creative task. Game mechanics should redirect attention away from judging the creative artifact and toward alternative goals such as participation, progression, social interaction, or strategy; time pressure alone is insufficient if success remains defined by producing the best-looking artifact.}

Moreover, evaluative off-centering involves a tradeoff between art quality and the volume of art created. Because critique can reintroduce self-evaluation, off-centering typically limits in-process feedback and de-emphasizes refinement, making it a poorer fit for larger and higher-stakes work that requires iterative critique.

Finally, evaluative off-centering is a reduction strategy: it may lessen momentary self-criticism but is unlikely to eliminate it, especially for creators with deep self-critical patterns.

\subsubsection{\new{Study Design}}
Our four-day between-subjects study ($N=38$) supports short-term effects but cannot establish durability or longer-term changes in self-criticism beyond the activity. Intermittent bugs reduced enjoyment on affected days and may have limited the effects. Pairings were between strangers rather than friends or a community, and partner engagement varied. Finally, our mechanics intentionally simplified gameplay and decoupled win/loss from art quality; different genres or tighter coupling may yield different effects.

\new{A core limitation of our evaluation is that ACG bundled multiple design features (e.g., social interaction, gameplay, art-as-a-ticket mechanism) relative to the control condition. While it is commonplace for new interaction systems to often integrate many co-occurring features~\cite{olsen2007evaluating}, we cannot causally isolate the effect of any single element or determine whether the observed outcomes arose from one component or their combination. Nonetheless, our qualitative data suggest social interaction may have contributed to enjoyment ratings, while treating art as a ticket to play rather an artifact to judge may have contributed to pride. Future work should disentangle which specific manipulations accounted for the outcomes.}

\new{A potential limitation in our study design was using art-history documentary videos as a filler activity in the control condition. Because these videos followed the illustration task but preceded the daily survey, they may have influenced participant ratings independently of the creative process itself. Exposure to well-known artworks could have prompted comparison with participants’ own sketches and affected pride ratings. However, qualitatively, no participants reported feeling discouraged or negatively comparing their own sketches to the video material. Several instead described the videos as ``interesting'' or ``enjoyable'', suggesting that the filler activity may have positively influenced enjoyment ratings independent of the illustration task. We observed no condition difference in excitement, making this concern most relevant to pride and enjoyment. We therefore ask readers to take this possible limitation into account when considering our quantitative results.}

\new{A further measurement limitation is that we measured a subset of three positive affect states rather than administering the full PANAS scale, and therefore do not comprehensively capture positive affect.}

\label{sec:discussion}

\section{Conclusion}
\newcc{We proposed \emph{evaluative off-centering}: a mechanism that redirects attention away from a self-evaluative task by embedding it within an intrinsically motivating surrounding activity. We instantiated this in Art Card Game (ACG), where time-bounded illustrations become immediately usable assets in head-to-head play. In a four-day randomized study with illustrators (\mbox{$N{=}38$}), ACG outperformed an active control with identical drawing constraints, increasing pride in produced artwork and enjoyment. These findings suggest evaluative off-centering as a promising design mechanism for creativity support tools that aim to reduce momentary self-criticism while keeping creators making.}

\bibliographystyle{ACM-Reference-Format}
\bibliography{sample-base}

\end{document}

%% file: likert-by-condition.tex
\begin{figure}[t]
  \centering
  \includegraphics[width=1.0\linewidth]{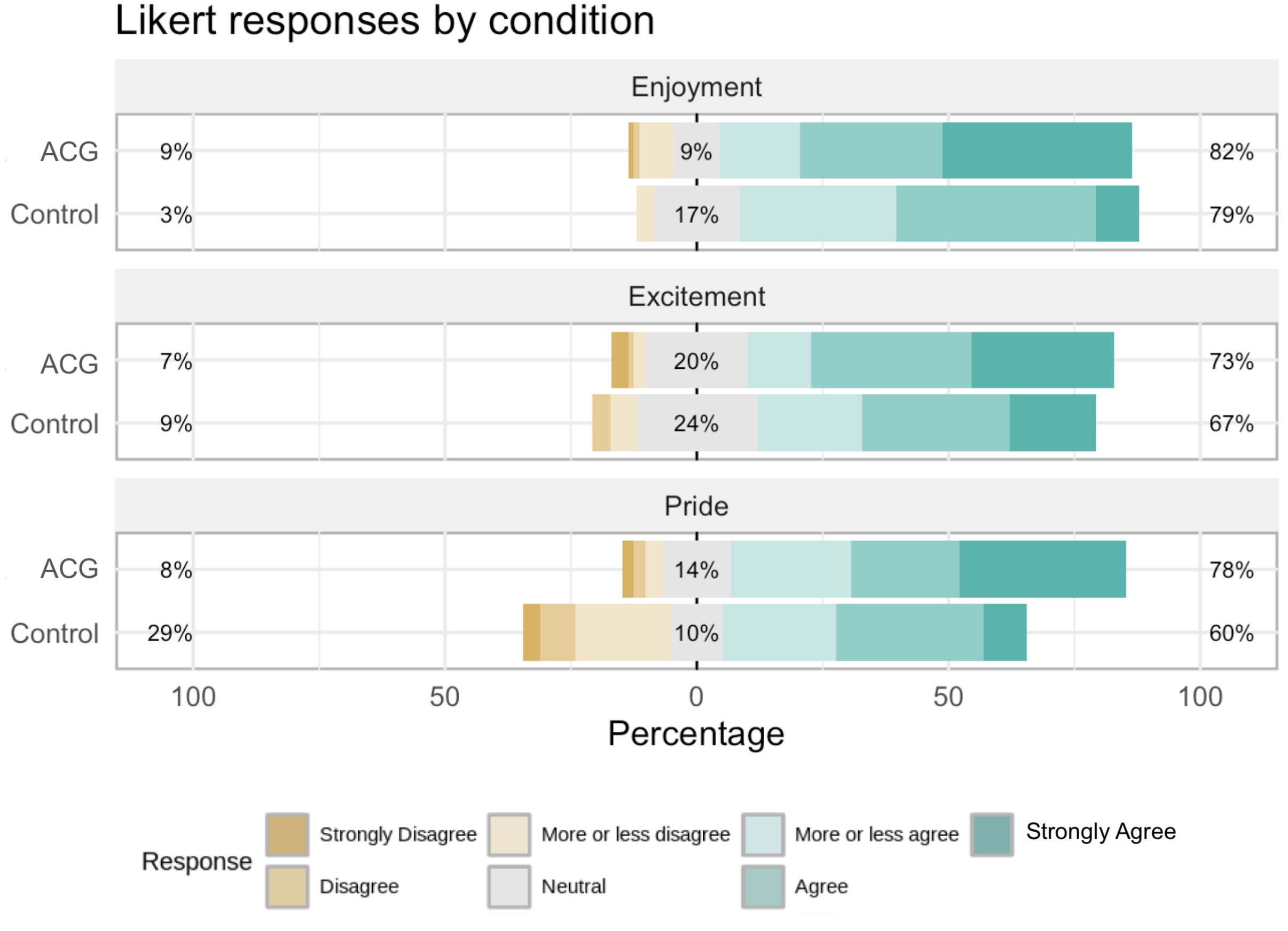}
  \caption{Descriptive distributions of 1–7 Likert responses (Pride, Enjoyment, Excitement) by condition (ACG vs.\ Control). Bars show the proportion at each category (stacked to 100\%). \new{The three percentages displayed on each bar represent the aggregate totals for all 'Disagree' categories (1–3), 'Neutral' responses (4), and all 'Agree' categories (5–7).} Art Card Game shows a visible rightward shift for Pride, a smaller shift for Enjoyment, and similar distributions for Excitement. Enjoyment has a noticeably higher proportion of "Strongly Agree" ratings in ACG compared to the control.}
  \Description{A stacked horizontal bar chart showing Likert response distributions for Enjoyment/Excitement/Pride across the 2 conditions. Each row contains two stacked bars (one per condition) that sum to 100\%, with segments representing the proportions selecting each response category: Strongly Disagree, Disagree, More or less disagree, Neutral, More or less agree, Agree, and Strongly Agree. The first row depicts the Enjoyment scoring, in which case for ACG, 9\% fall in the three disagreement categories combined (Strongly Disagree, Disagree, More or less disagree), 82\% fall in the agreement categories (More or less agree, Agree, Strongly Agree), and the remaining 9\% in neutral. For Control, 3\% fall in disagreement, 79\% in agreement, and 17\% in neutral. The second row depicts Excitement scores. For ACG, 7\% fall in disagreement, 73\% in agreement, and 20\% in neutral. For Control, 9\% fall in disagreement, 67\% in agreement, and 24\% in neutral. The last row depicts Pride scores. For ACG, 8\% fall in disagreement, 78\% in agreement, and 14\% in neutral. For Control, 29\% fall in disagreement, 60\% in agreement, and 10\% in neutral.}
  \label{fig:likert-overall}
\end{figure}

%% file: forest-plot.tex
\begin{figure}[t]
  \centering
  \includegraphics[width=1.0\linewidth]{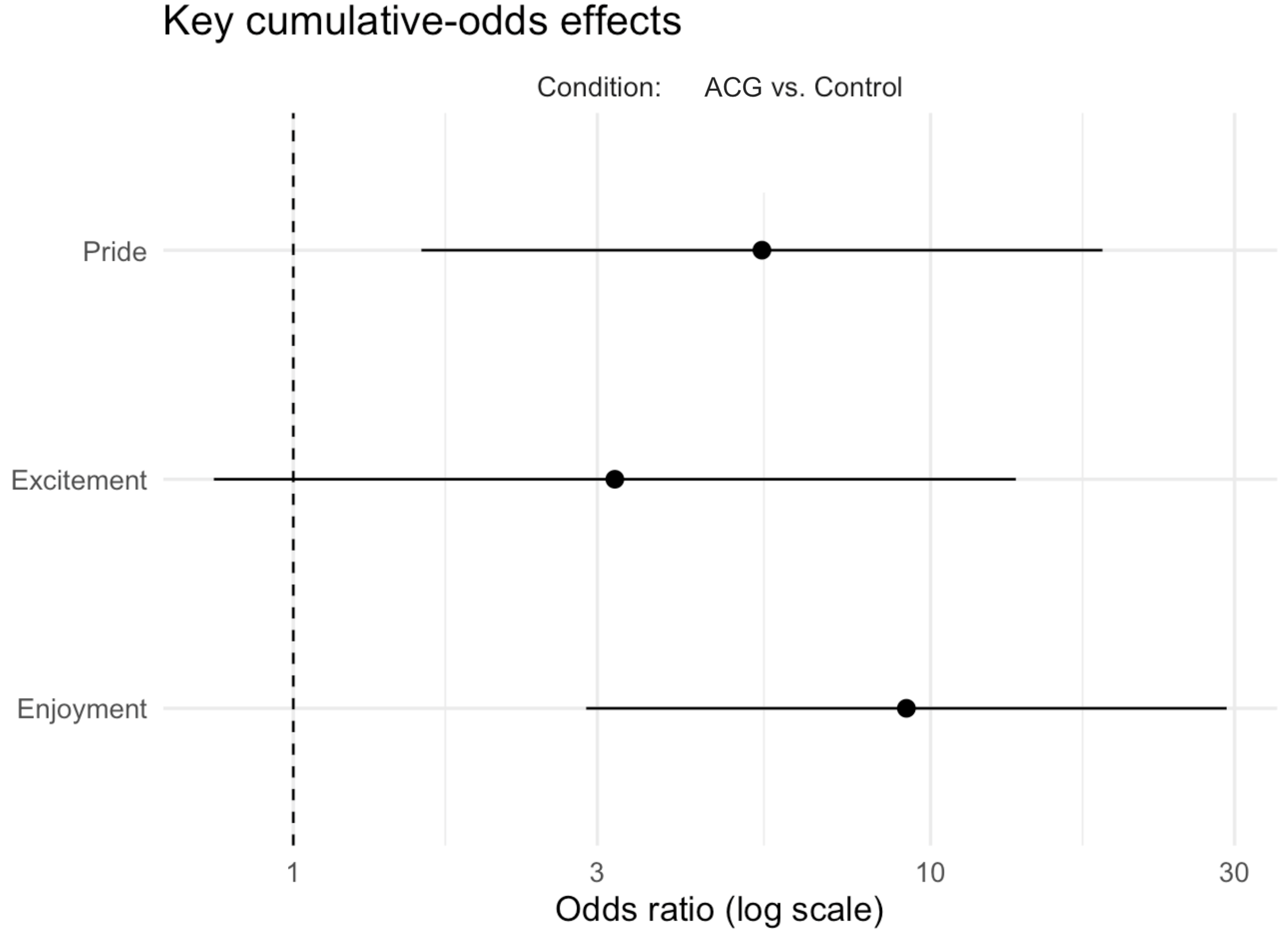}
  \caption{Cumulative link mixed model (ordered logistic) fixed effects. Points show odds ratios (OR) with 95\% CIs for Condition (ACG vs.\ Control), Study Day, and the bug indicator (ACG only), estimated separately for Pride, Enjoyment, and Excitement. OR$>$1 indicates higher odds of a higher category.}
  \Description{A horizontal forest plot showing odds ratios for measures: Pride, Enjoyment, and Excitement. Each measure displays an odds ratio point estimate with a horizontal confidence interval bar. The plot begins at an odds ratio value of 1. The interval bars for Pride and Enjoyment are largely to the right of 1, with Enjoyment further right than Pride, indicating higher odds of reporting a higher Likert category in the ACG condition. The Excitement interval bar is shifted more left, starting behind 1, indicating little or no effect. Exact odds ratios and confidence intervals are reported in the manuscript.}
  \label{fig:forest}
\end{figure}

%% file: clmm_three_models_with_day_stars_from_summary.tex
\begin{table*}[t]
\centering
\begin{tabular}{l p{2.2cm} p{2.2cm} p{2.2cm}}
\toprule
 & \centering Enjoyment & \centering Pride & \centering Excitement \tabularnewline
\cmidrule(lr){2-2}\cmidrule(lr){3-3}\cmidrule(lr){4-4}
Condition: ACG (baseline: Control) & \centering 2.215$^{***}$ \par (0.590) & \centering 1.693$^{**}$ \par (0.628) & \centering 1.162 \par (0.739) \tabularnewline
\addlinespace[4pt]
Experienced Bug (nested in ACG) & \centering -3.706$^{***}$ \par (0.600) & \centering -0.574 \par (0.501) & \centering -1.213$^{*}$ \par (0.515) \tabularnewline
\addlinespace[4pt]
Study day: 2 & \centering -0.203 \par (0.445) & \centering 0.896$^{*}$ \par (0.441) & \centering -0.027 \par (0.436) \tabularnewline
\addlinespace[4pt]
Study day: 3 & \centering -0.028 \par (0.451) & \centering 1.052$^{*}$ \par (0.440) & \centering -1.036$^{*}$ \par (0.444) \tabularnewline
\addlinespace[4pt]
Study Day: 4 & \centering -0.267 \par (0.473) & \centering 1.073$^{*}$ \par (0.471) & \centering -1.054$^{*}$ \par (0.504) \tabularnewline
\midrule
AIC    & \centering 409.600 & \centering 487.517 & \centering 439.802 \tabularnewline
LogLik & \centering -192.800 & \centering -231.758 & \centering -207.901 \tabularnewline
N      & \centering 146 & \centering 146 & \centering 146 \tabularnewline
\bottomrule
\end{tabular}
\caption{Art Card Game increased Enjoyment and Pride in participants' artmaking, compared to the control condition. The model controls for whether participants experienced a game-breaking bug, which negatively impacted enjoyment and excitement. Ordinal mixed-effects with study-day fixed effects (baseline = Day 1). Cumulative-logit coefficients; standard errors in parentheses. * $p<0.05$, ** $p<0.01$, *** $p<0.001$.}
\label{tab:clmm_three_stars}
\end{table*}

%% file: sample-base.bib
@String{Computing = "Computing" }

@String{Computer = "{IEEE} Computer" }

@String{Academic = "Academic Press" }

@String{Springer = "Springer-Verlag" }

@online{ToTD,
  author =    {SEGA},
  title =  {The Typing of The Dead: Overkill},
  year = 2013,
  url =    {https://store.steampowered.com/app/246580/The_Typing_of_The_Dead_Overkill/
},
  organization = {SEGA},
  lastaccessed = {January 27, 2026}
}

@online{hrm,
  author =    {Tomorrow Corporation},
  title =  {Human Resource Machine},
  year = 2015,
  organization = {Tomorrow Corporation},
  url =    {https://store.steampowered.com/app/375820/Human_Resource_Machine/},
  lastaccessed = {January 27, 2026}
}

@inproceedings{beyondproductivity,
author = {Rhys Cox, Samuel and B\o{}jer Djern\ae{}s, Helena and van Berkel, Niels},
title = {Beyond Productivity: Rethinking the Impact of Creativity Support Tools},
year = {2025},
isbn = {9798400712890},
publisher = {Association for Computing Machinery},
address = {New York, NY, USA},
url = {https://doi.org/10.1145/3698061.3726924},
doi = {10.1145/3698061.3726924},
booktitle = {Proceedings of the 2025 Conference on Creativity and Cognition},
pages = {735–749},
numpages = {15},
location = {
},
series = {C\&C '25}
}

@online{ZRun,
  author =    {Alderman, Naomi},
  title =  {ZRX - Transform your workout into adventure!},
  year = 2012,
  organization = {Six to Start},
  url = {https://zrx.app/},
  lastaccessed = {January 27, 2026}
}

@online{InkTober,
  title =        "InkTober",
  url =          "https://inktober.com/",
  lastaccessed = "September 3, 2024",
year=2013,
author="InkTober",
organization="InkTober"
}

@online{GRTPHNE,
  title =        "Gartic Phone",
  url =          "https://garticphone.com/",
  lastaccessed = "September 3, 2024",
year=2020,
author="Gartic Phone",
  organization="Onrizon Social Games"
}

@article{hwang2023effect,
  title={Effect of Toothbrushing Application for Kids on Dental Plaque Removal and Interest in Toothbrushing of Preschool Children},
  author={Hwang, Chae-Ha and Song, Hyeon-Ju and Jung, Min-Ji and Choi, Yeon-Jae and Hwang, Young Sun},
  journal={Journal of dental hygiene science},
  volume={23},
  number={3},
  pages={208--215},
  year={2023},
  publisher={The Korean Society of Dental Hygiene Science}
}

@inproceedings{leeko2011,
author = {Lee, Michael J. and Ko, Amy J.},
title = {Personifying programming tool feedback improves novice programmers' learning},
year = {2011},
isbn = {9781450308298},
publisher = {Association for Computing Machinery},
address = {New York, NY, USA},
url = {https://doi.org/10.1145/2016911.2016934},
doi = {10.1145/2016911.2016934},
booktitle = {Proceedings of the Seventh International Workshop on Computing Education Research},
pages = {109–116},
numpages = {8},
location = {Providence, Rhode Island, USA},
series = {ICER '11}
}

@book{lefcourt1991locus,
  title={Locus of control.},
  author={Lefcourt, Herbert M},
  year={1991},
  address = {San Diego, CA, USA},
  publisher={Academic Press}
}

@article{crawford1984viktor,
  title={Viktor Shklovskij: Diff{\'e}rance in Defamiliarization},
  author={Crawford, Lawrence},
  journal={Comparative Literature},
  pages={209--219},
  year={1984},
  volume = {36},
  number = {3},
  publisher={JSTOR}
}

@inproceedings{noguchi1999material,
  title={How do material constraints affect design creativity?},
  author={Noguchi, Hisataka},
  booktitle={Proceedings of the 3rd Conference on Creativity \& Cognition},
  pages={82--87},
    publisher = {Association for Computing Machinery},
  address = {New York, NY, USA},
  year={1999}
}

@inproceedings{carlson2013designing,
  title={Designing interaction for designers: defamiliarization in user's creative decision-making},
  author={Carlson, Kristin and Schiphorst, Thecla},
  booktitle={Proceedings of the 9th ACM Conference on Creativity \& Cognition},
  pages={300--303},
  year={2013},
  publisher = {Association for Computing Machinery},
  address = {New York, NY, USA},
}

@online{redditPerfection1,
  title =        "How do you stop with the perfectionism and self-sabotage?",
  url =          "https://www.reddit.com/r/ArtistLounge/comments/k7w259/how_do_you_stop_with_the_perfectionism_and/?rdt=34967",
  lastaccessed = "September 3, 2024",
    author="Reddit",
    year=2005,
organization="Reddit"
}

@online{r-artistlounge,
  title =        "Artist Lounge",
  url =          "https://www.reddit.com/r/ArtistLounge/",
  lastaccessed = "September 9, 2025",
    organization="Reddit",
author="Reddit",
    year=2013,
}

@misc{adamduff2020,
  author       = {Adam Duff LUCIDPIXUL}, 
  title        = {Advice For Artists Who Are Too Hard On Themselves},
  year         = {2020},
  month        = {March},
  note         = {{YouTube video. Available at: \url{https://www.youtube.com/watch?v=w6StfT6_a7g} (accessed 2025-09-10)}}
}

@misc{chiu2016,
  author       = {Bobby Chiu}, 
  title        = {Artist Self Doubt},
  year         = {2016},
  month        = {April},
  note         = {{YouTube video. Available at:https://www.youtube.com/watch?v=huwOsyPseII} (accessed 2025-09-10)}}

@article{KochJulieM2024FASM,
title = {Fine Arts Students: Mental Health, Stress, and Time on Academic Work},
author = {Lee, Fallyn M. and Koch, Julie M. and Ramakrishnan, Nikita},
doi = {10.1080/87568225.2023.2175755},
issn = {8756-8225},
journal = {Journal of college student psychotherapy},
language = {eng},
number = {2},
pages = {259 - 274},
publisher = {Routledge},
volume = {38},
year = {2024},
}

@phdthesis{sternam2022,
  title={Process-Sensitive Creativity Support Tools},
  author={Sterman, Sarah Gimbert},
  year={2022},
  school={University of California, Berkeley}
}

@inproceedings{kim2017mosaic,
author = {Kim, Joy and Agrawala, Maneesh and Bernstein, Michael S.},
title = {Mosaic: Designing Online Creative Communities for Sharing Works-in-Progress},
year = {2017},
isbn = {9781450343350},
publisher = {Association for Computing Machinery},
address = {New York, NY, USA},
url = {https://doi.org/10.1145/2998181.2998195},
doi = {10.1145/2998181.2998195},
booktitle = {Proceedings of the 2017 ACM Conference on Computer Supported Cooperative Work and Social Computing},
pages = {246–258},
numpages = {13},
location = {Portland, Oregon, USA},
series = {CSCW '17}
}

@article{SION2023102060,
title = {Self-correction, digital art making and stress reduction},
journal = {The Arts in Psychotherapy},
volume = {85},
pages = {102060},
year = {2023},
issn = {0197-4556},
doi = {https://doi.org/10.1016/j.aip.2023.102060},
url = {https://www.sciencedirect.com/science/article/pii/S0197455623000679},
author = {A. Sion and J. Czamanski-Cohen and O.C. Halbrecht-Shaked and G. Galili and J. Cwikel},
}

@inproceedings{belakova2021,
author = {Belakova, Jekaterina and Mackay, Wendy E.},
title = {SonAmi: A Tangible Creativity Support Tool for Productive Procrastination},
year = {2021},
isbn = {9781450383769},
publisher = {Association for Computing Machinery},
booktitle = {Proceedings of the 13th Conference on Creativity and Cognition},
address = {New York, NY, USA},
url = {https://doi.org/10.1145/3450741.3465250},
doi = {10.1145/3450741.3465250},
articleno = {7},
numpages = {10},
location = {Virtual Event, Italy},
series = {C\&C '21}
}

@phdthesis{standish2022,
  title={The impact of self-criticism throughout the creative process},
  author={Standish, Colleen},
  year={2022},
  school={The University of Oklahoma}
}

@online{thumbnailing2023,
  title =        "Guide to Thumbnail Sketches",
  url =          "https://artprof.org/learn/tutorials-media/drawing/thumbnail-sketches/",
  lastaccessed = "September 3, 2024",
    author="Art Prof",
    year=2023,
  organization="Art Prof"
}

@inproceedings{williford2017zensketch,
    author = {Williford, Blake and Runyon, Matthew and Malla, Adil Hamid and Li, Wayne and Linsey, Julie and Hammond, Tracy},
    title = {ZenSketch: A Sketch-based Game For Improving Line Work},
    year = {2017},
    isbn = {9781450351119},
    publisher = {Association for Computing Machinery},
    address = {New York, NY, USA},
    url = {https://doi.org/10.1145/3130859.3130861},
    doi = {10.1145/3130859.3130861},
    booktitle = {Extended Abstracts Publication of the Annual Symposium on Computer-Human Interaction in Play},
    pages = {591–598},
    numpages = {8},
    location = {Amsterdam, The Netherlands},
    series = {CHI PLAY '17 Extended Abstracts}
}

@article{dow2010,
author = {Dow, Steven P. and Glassco, Alana and Kass, Jonathan and Schwarz, Melissa and Schwartz, Daniel L. and Klemmer, Scott R.},
title = {Parallel prototyping leads to better design results, more divergence, and increased self-efficacy},
year = {2011},
issue_date = {December 2010},
publisher = {Association for Computing Machinery},
address = {New York, NY, USA},
volume = {17},
number = {4},
issn = {1073-0516},
url = {https://doi.org/10.1145/1879831.1879836},
doi = {10.1145/1879831.1879836},
journal = {ACM Trans. Comput.-Hum. Interact.},
month = dec,
articleno = {18},
numpages = {24}
}

@article{GLAZIEWICZ2024101692,
title = {When the creative well dries up–burnout syndrome and art block in artists’ sample},
journal = {Thinking Skills and Creativity},
volume = {54},
pages = {101692},
year = {2024},
issn = {1871-1871},
doi = {https://doi.org/10.1016/j.tsc.2024.101692},
url = {https://www.sciencedirect.com/science/article/pii/S187118712400230X},
author = {Karolina Głaziewicz and Krystyna Golonka},
}

@article{BASAK20125010,
title = {Perfectionist Attitudes of Artistically Talented Students in the Art Classroom},
journal = {Procedia - Social and Behavioral Sciences},
volume = {46},
pages = {5010-5014},
year = {2012},
note = {4th WORLD CONFERENCE ON EDUCATIONAL SCIENCES (WCES-2012) 02-05 February 2012 Barcelona, Spain},
issn = {1877-0428},
doi = {https://doi.org/10.1016/j.sbspro.2012.06.377},
url = {https://www.sciencedirect.com/science/article/pii/S1877042812021131},
author = {Rasim Basak},
}

@article{Germer2013SelfcompassionIC,
  title={Self-compassion in clinical practice.},
  author={Christopher K. Germer and Kristin D. Neff},
  journal={Journal of clinical psychology},
  year={2013},
  volume={69 8},
  pages={
          856-67
        },
  url={https://api.semanticscholar.org/CorpusID:2799698}
}

@article{ong2019randomized,
  title={A randomized controlled trial of acceptance and commitment therapy for clinical perfectionism},
  author={Ong, Clarissa W and Lee, Eric B and Krafft, Jennifer and Terry, Carina L and Barrett, Tyson S and Levin, Michael E and Twohig, Michael P},
  journal={Journal of Obsessive-Compulsive and Related Disorders},
  volume={22},
  pages={100444},
  year={2019},
  publisher={Elsevier}
}

@article{galloway2022efficacy,
  title={The efficacy of randomised controlled trials of cognitive behaviour therapy for perfectionism: A systematic review and meta-analysis},
  author={Galloway, Ricky and Watson, Hunna and Greene, Danyelle and Shafran, Roz and Egan, Sarah J},
  journal={Cognitive Behaviour Therapy},
  volume={51},
  number={2},
  pages={170--184},
  year={2022},
  publisher={Taylor \& Francis}
}

@article{zabelina2010don,
  title={Don't be so hard on yourself: Self-compassion facilitates creative originality among self-judgmental individuals},
  author={Zabelina, Darya L and Robinson, Michael D},
  journal={Creativity Research Journal},
  volume={22},
  number={3},
  pages={288--293},
  year={2010},
  publisher={Taylor \& Francis}
}

@article{shahar2012pilot,
  title={A pilot investigation of emotion-focused two-chair dialogue intervention for self-criticism},
  author={Shahar, Ben and Carlin, Erica R and Engle, David E and Hegde, Jayanta and Szepsenwol, Ohad and Arkowitz, Hal},
  journal={Clinical psychology \& psychotherapy},
  volume={19},
  number={6},
  pages={496--507},
  year={2012},
  publisher={Wiley Online Library}
}

@article{carson2011,
author = {Carson, Shelley H.},
title ={Creativity and Psychopathology: A Shared Vulnerability Model},
journal = {The Canadian Journal of Psychiatry},
volume = {56},
number = {3},
pages = {144-153},
year = {2011},
doi = {10.1177/070674371105600304}
}

@book{pressfield2002,
  title={The War of Art: Break Through the Blocks and Win Your Inner Creative Battles},
  author={Pressfield, Steven},
  year={2002},
  publisher={Black Irish Entertainment LLC},
  address={USA}
}

@book{whatisillustration,
  title={What is Illustration? (Essential Design Handbook)},
  author={Zeegen, Lawrence},
  year={2009},
  publisher={Rockport Publishers},
  address={USA}
}

@online{nea2024arts,
  author  = {National Endowment for the Arts},
  title   = {New Research Explores Arts Engagement and Social Connectedness},
  url     = {https://www.arts.gov/news/press-releases/2024/new-research-explores-arts-engagement-and-social-connectedness},
  year = {2024},
  organization = {National Endowment for the Arts}
}

@article{watson1988development,
  title={Development and validation of brief measures of positive and negative affect: the PANAS scales.},
  author={Watson, David and Clark, Lee Anna and Tellegen, Auke},
  journal={Journal of personality and social psychology},
  volume={54},
  number={6},
  pages={1063},
  year={1988},
  publisher={American Psychological Association}
}

@inproceedings{olsen2007evaluating,
  title={Evaluating user interface systems research},
  author={Olsen Jr, Dan R},
  booktitle={Proceedings of the 20th annual ACM symposium on User interface software and technology},
  pages={251--258},
  year={2007},
  publisher = {Association for Computing Machinery},
  address={USA}
}

@article{gilbert2004criticizing,
  title={Criticizing and reassuring oneself: An exploration of forms, styles and reasons in female students},
  author={Gilbert, Paul and Clarke, Marlea and Hempel, Susanne and Miles, Jeremy NV and Irons, Chris},
  journal={British Journal of Clinical Psychology},
  volume={43},
  number={1},
  pages={31--50},
  year={2004},
  publisher={Wiley Online Library}
}

@article{zuroff2016conceptualizing,
  title={Conceptualizing and measuring self-criticism as both a personality trait and a personality state},
  author={Zuroff, David C and Sadikaj, Gentiana and Kelly, Allison C and Leybman, Michelle J},
  journal={Journal of personality assessment},
  volume={98},
  number={1},
  pages={14--21},
  year={2016},
  publisher={Taylor \& Francis}
}

@Inbook{chen2009,
author="Chen, Peayton
and Kuo, Rita
and Chang, Maiga
and Heh, Jia-Sheng",
title="Designing a Trading Card Game as Educational Reward System to Improve Students' Learning Motivations",
bookTitle="Transactions on Edutainment III",
year="2009",
publisher="Springer Berlin Heidelberg",
address="Berlin, Heidelberg",
pages="116--128",
isbn="978-3-642-11245-4",
doi="10.1007/978-3-642-11245-4_11",
url="https://doi.org/10.1007/978-3-642-11245-4_11"
}

@online{discord-doxxing-policy,
  author  = {{Discord, Inc.}},
  title   = {Doxxing Policy Explainer},
  url     = {https://discord.com/safety/doxxing-policy-explainer},
  urldate = {2025-11-27},
  year = {2025},
  organization = {Discord, Inc.}
}

@article{carron1996social,
  author  = {Carron, Albert V. and Hausenblas, Heather A. and Mack, Diane},
  title   = {Social Influence and Exercise: A Meta-Analysis},
  journal = {Journal of Sport and Exercise Psychology},
  year    = {1996},
  volume  = {18},
  number  = {1},
  pages   = {1--16}
}

@article{neff2023review,
   author = "Neff, Kristin D.",
   title = "Self-Compassion: Theory, Method, Research, and Intervention", 
   journal= "Annual Review of Psychology",
   year = "2023",
   volume = "74",
   number = "Volume 74, 2023",
   pages = "193-218",
   doi = "https://doi.org/10.1146/annurev-psych-032420-031047",
   url = "https://www.annualreviews.org/content/journals/10.1146/annurev-psych-032420-031047",
   publisher = "Annual Reviews",
   issn = "1545-2085",
   type = "Journal Article"
}

@article{lin2024socialsupport,
author = {Lin, Hao and Chen, Haidong and Liu, Qingzao and Xu, Jie and Li, Shan},
year = {2024},
month = {01},
pages = {},
title = {A meta-analysis of the relationship between social support and physical activity in adolescents: the mediating role of self-efficacy},
volume = {14},
journal = {Frontiers in Psychology},
doi = {10.3389/fpsyg.2023.1305425}
}

@book{character2020,
  title={Fundamentals of Character Design: How to Create Engaging Characters for Illustration, Animation \& Visual Development},
  author={Bishop, Randy and Boo, Sweeney and Crus Meybis R. and Gadea, Luis},
  year={2020},
  publisher={3dtotal Publishing},
  address={USA}
}

@online{af,
  title = "Welcome to Art Fight! The annual art trading game, starting on July 1st!",
  author =        "Art Fight",
  url =          "https://artfight.net/",
  year = 2015,
  lastaccessed = "September 3, 2024",
  organization = "Art Fight"
}

@book{isbister2016games,
  title={How games move us: Emotion by design},
  author={Isbister, Katherine},
  year={2016},
  publisher={Mit Press},
  address={USA}
}

@article{kaufman2012changing,
  title={Changing beliefs and behavior through experience-taking.},
  author={Kaufman, Geoff F and Libby, Lisa K},
  journal={Journal of personality and social psychology},
  volume={103},
  number={1},
  pages={1},
  year={2012},
  publisher={American Psychological Association}
}

@article{isbister2016connecting,
  title={Connecting through play},
  author={Isbister, Katherine},
  journal={Interactions},
  volume={23},
  number={4},
  pages={26--33},
  year={2016},
  publisher={ACM New York, NY, USA}
}

@article{villani2018videogames,
  title={Videogames for emotion regulation: a systematic review},
  author={Villani, Daniela and Carissoli, Claudia and Triberti, Stefano and Marchetti, Antonella and Gilli, Gabriella and Riva, Giuseppe},
  journal={Games for health journal},
  volume={7},
  number={2},
  pages={85--99},
  year={2018},
  publisher={Mary Ann Liebert, Inc. 140 Huguenot Street, 3rd Floor New Rochelle, NY 10801 USA}
}

@article{hantsbarger2022alienated,
  title={Alienated serendipity and reflective failure: Exploring queer game mechanics and queerness in games via queer temporality},
  author={Hantsbarger, Matthew and Troiano, Giovanni Maria and To, Alexandra and Harteveld, Casper},
  journal={Proceedings of the ACM on Human-Computer Interaction},
  volume={6},
  number={CHI PLAY},
  pages={1--27},
  year={2022},
  publisher={ACM New York, NY, USA}
}

@article{to2018character,
author = {To, Alexandra and McDonald, Joselyn and Holmes, Jarrek and Kaufman, Geoff and Hammer, Jessica},
year = {2018},
month = {10},
pages = {},
title = {Character Diversity in Digital and Non-Digital Games},
volume = {4},
journal = {Transactions of the Digital Games Research Association},
doi = {10.26503/todigra.v4i1.84}
}

@phdthesis{bryan2025playing,
  title={Playing Karaoke: A Lived Experience of Play as Method},
  author={Bryan, Jeffrey S},
  year={2025},
  school={University of California, Irvine}
}

@ARTICLE{2009JSEdT,
       author = {{Chase}, Catherine C. and {Chin}, Doris B. and {Oppezzo}, Marily A. and {Schwartz}, Daniel L.},
        title = "{Teachable Agents and the Prot{\'e}g{\'e} Effect: Increasing the Effort Towards Learning}",
      journal = {Journal of Science Education and Technology},
         year = 2009,
        month = aug,
       volume = {18},
       number = {4},
        pages = {334-352},
          doi = {10.1007/s10956-009-9180-4},
       adsurl = {https://ui.adsabs.harvard.edu/abs/2009JSEdT..18..334C}
}

@article{hart1989tie,
  title={Tie measurement of social physique anxiety},
  author={Hart, Elizabeth A and Leary, Mark R and Rejeski, W Jack},
  journal={Journal of Sport and exercise Psychology},
  volume={11},
  number={1},
  pages={94--104},
  year={1989},
  publisher={Human Kinetics, Inc.}
}

@ARTICLE{4569871,
  author={Connolly, Cornelia and Murphy, Eamonn and Moore, Sarah},
  journal={IEEE Transactions on Education}, 
  title={Programming Anxiety Amongst Computing Students—A Key in the Retention Debate?}, 
  year={2009},
  volume={52},
  number={1},
  pages={52-56},
  doi={10.1109/TE.2008.917193}
}

@article{reeves1996media,
  title={The media equation: How people treat computers, television, and new media like real people},
  author={Reeves, Byron and Nass, Clifford},
  journal={Cambridge, UK},
  volume={10},
  number={10},
  pages={19--36},
  year={1996}
}

@inproceedings{murnane2020designing,
  title={Designing ambient narrative-based interfaces to reflect and motivate physical activity},
  author={Murnane, Elizabeth L and Jiang, Xin and Kong, Anna and Park, Michelle and Shi, Weili and Soohoo, Connor and Vink, Luke and Xia, Iris and Yu, Xin and Yang-Sammataro, John and others},
  booktitle={Proceedings of the 2020 CHI Conference on Human Factors in Computing Systems},
  publisher = {Association for Computing Machinery},
  address = {New York, NY, USA},
  pages={1--14},
  year={2020}
}
